\pdfoutput=1 

\PassOptionsToPackage{bookmarks=true,breaklinks=true,letterpaper=true,colorlinks,linkcolor=black,citecolor=black,urlcolor=black}{hyperref}
\documentclass[sigconf]{acmart}

\copyrightyear{2021}
\acmYear{2021}
\setcopyright{acmlicensed}\acmConference[MICRO '21]{54th Annual IEEE/ACM International Symposium on Microarchitecture}{October 18--22, 2021}{Virtual Event, Greece}
\acmBooktitle{Proceedings of the 54th Annual IEEE/ACM International Symposium on Microarchitecture (MICRO '21), October 18--22, 2021, Virtual Event, Greece}
\acmPrice{15.00}
\acmDOI{10.1145/3466752.3480061}
\acmISBN{978-1-4503-8557-2/21/10}

\DeclareMathAlphabet{\mathcal}{OMS}{cmsy}{m}{n}
\usepackage[tt=false]{libertine}

\usepackage{array}

\newcommand{\ignore}[1]{}
\usepackage{fancyhdr}
\usepackage{datetime}
\usepackage[bookmarks=true,breaklinks=true,letterpaper=true,colorlinks,linkcolor=black,citecolor=black,urlcolor=black]{hyperref}
\usepackage{multirow}
\usepackage{multicol}
\usepackage{float}
\usepackage[shortlabels]{enumitem}
\usepackage[nomessages]{fp}
\usepackage{dblfloatfix}    

\usepackage{bm}
\usepackage{tikz}
\usepackage{atbegshi}
\usepackage{xcolor}
\usepackage[font={small,bf}]{subcaption}

\usepackage[font={small,bf}]{caption}
\setlist[itemize]{leftmargin=*}

\usepackage{amsmath}
\usepackage{cleveref}

\crefformat{section}{\S#2#1#3}
\crefformat{subsection}{\S#2#1#3}
\crefformat{subsubsection}{\S#2#1#3}
\crefformat{appendix}{\S#2#1#3}

\setlist[itemize]{noitemsep,itemsep=0pt,parsep=0pt,topsep=0pt,partopsep=0pt,leftmargin=1.5em}
\setlist[enumerate]{noitemsep,itemsep=0pt,parsep=0pt,topsep=0pt,partopsep=0pt,leftmargin=1.5em}

\newcolumntype{L}[1]{>{\raggedright\let\newline\\\arraybackslash\hspace{0pt}}m{#1}}
\newcolumntype{C}[1]{>{\centering\let\newline\\\arraybackslash\hspace{0pt}}m{#1}}
\newcolumntype{R}[1]{>{\raggedleft\let\newline\\\arraybackslash\hspace{0pt}}m{#1}}

\definecolor{amber}{rgb}{1.0, 0.49, 0.0}
\definecolor{darkamber}{rgb}{0.9, 0.49, 0.0}
\definecolor{extradarkamber}{rgb}{0.55, 0.295, 0.0}
\definecolor{darkgreen}{rgb}{0.0, 0.2, 0.13}
\definecolor{darkbyzantium}{rgb}{0.36, 0.22, 0.33}
\definecolor{darkseagreen}{rgb}{0.56, 0.74, 0.56}
\definecolor{darkspringgreen}{rgb}{0.09, 0.45, 0.27}
\definecolor{dollarbill}{rgb}{0.52, 0.73, 0.4}
\definecolor{darkcerulean}{rgb}{0.03, 0.27, 0.49}
\definecolor{poop}{rgb}{0.75, 0.33, 0.0}

\DeclareRobustCommand*{\mydarkamber}{\textcolor{darkamber}}

\def\mechanism/{XYZ}

\newif\ifcameraready
\camerareadytrue

\ifcameraready
  \usepackage[disable]{todonotes}

  \newcommand{\mpi}[1]{#1}
  \newcommand{\mpj}[1]{#1}
  \newcommand{\mpk}[1]{#1}
  
  \newcommand{\mpm}[1]{#1}
  \newcommand{\mpn}[1]{#1}
  \newcommand{\mpp}[1]{#1}
  \newcommand{\mpq}[1]{#1}
  \newcommand{\mpr}[1]{#1}
  \newcommand{\mps}[1]{#1}
  \newcommand{\mpu}[1]{#1}
  \newcommand{\mpv}[1]{#1}

  \newcommand{\gfi}[1]{#1}
  \newcommand{\gfii}[1]{#1}

  \newcommand{\rev}[1]{#1}
\else
  \usepackage[textsize=scriptsize]{todonotes}
  \usepackage{marginnote}
  \let\marginpar\marginnote

  \newcommand{\mpi}[1]{#1}
  \newcommand{\mpj}[1]{#1}
  \newcommand{\mpk}[1]{#1}
  
  \newcommand{\mpm}[1]{#1} 
  \newcommand{\mpn}[1]{#1} 
  \newcommand{\mpp}[1]{#1} 
  \newcommand{\mpq}[1]{#1} 
  \newcommand{\mpr}[1]{#1} 
  \newcommand{\mps}[1]{#1} 
  \newcommand{\mpu}[1]{#1} 
  \newcommand{\mpv}[1]{\mydarkamber{#1}}

  \newcommand{\gfi}[1]{#1} 
   \newcommand{\gfii}[1]{#1}

  \newcommand{\rev}[1]{#1}
  
\fi

\makeatletter
\newcommand*{\textoverline}[1]{$\overline{\raisebox{0pt}[0.85\height]{#1}}\m@th$}
\makeatother

\begin{document}

\title[HARP: Practically and Effectively Identifying Uncorrectable Errors\\in Memory Chips That Use On-Die Error-Correcting Codes]{HARP: Practically and Effectively\\Identifying Uncorrectable Errors in Memory Chips\\That Use On-Die Error-Correcting Codes}

\newcommand{\ethz}{{\large$^\dagger$}}
\newcommand{\cmu}{{\large$^\ddagger$}}
\newcommand{\scomma}{{\large$^,$}}

\author{Minesh Patel}
\affiliation{%
  \institution{ETH Z{\"u}rich}
  \country{}
}
\author{Geraldo F. Oliveira}
\affiliation{%
  \institution{ETH Z{\"u}rich}
  \country{}
}
\author{Onur Mutlu}
\affiliation{%
  \institution{ETH Z{\"u}rich}
  \country{}
}
\renewcommand{\shortauthors}{Patel, et al.}

\sloppypar
\begin{abstract}
Aggressive storage density scaling in modern main memories causes increasing
error rates that are addressed using error-mitigation techniques.
State-of-the-art techniques for addressing high error rates identify and repair
bits that are at risk of error from within the memory controller. Unfortunately,
modern main memory chips internally use on-die error correcting codes (on-die
ECC) that obfuscate the memory controller's view of errors, complicating the
process of identifying at-risk bits (i.e., error profiling).

To understand the problems that on-die ECC causes for error profiling, we
analytically study how on-die ECC changes the way that memory errors appear
outside of the memory chip (e.g., to the memory controller). We show that on-die
ECC introduces statistical dependence between errors in different bit positions,
raising three key challenges for practical and effective error profiling: on-die
ECC (1) exponentially increases the number of at-risk bits the profiler must
identify; (2) makes individual at-risk bits more difficult to identify; and (3)
interferes with commonly-used memory data patterns that are designed to make
at-risk bits easier to identify.

To address the three challenges, we introduce Hybrid Active-Reactive Profiling
(HARP), a new error profiling algorithm that rapidly achieves full coverage of
at-risk bits based on two key insights. First, errors that on-die ECC fails to
correct have two sources: (1) direct errors from raw bit errors in the data
portion of the ECC word and (2) indirect errors that on-die ECC introduces when
facing uncorrectable errors. Second, the maximum number of indirect errors that
can occur concurrently is limited to the correction capability of on-die ECC.
HARP's key idea is to first identify all bits at risk of direct errors using
existing profiling techniques with the help of small modifications to the on-die
ECC mechanism. Then, a secondary ECC within the memory controller with
correction capability equal to or greater than that of on-die ECC can safely
identify bits at-risk of indirect errors, if and when they fail. 

We evaluate HARP in simulation relative to two state-of-the-art baseline error
profiling algorithms. We show that HARP achieves full coverage of all at-risk
bits faster (e.g., 99th-percentile coverage 20.6\%/36.4\%/52.9\%/62.1\% faster,
on average, given 2/3/4/5 raw bit errors per ECC word) than the baseline
algorithms, which sometimes fail to achieve full coverage. We perform a case
study of how each profiler impacts the system's overall bit error rate (BER)
when using a repair mechanism to tolerate DRAM data-retention errors. We show
that HARP identifies all errors faster than the best-performing baseline
algorithm (e.g., by $3.7\times$ for a raw per-bit error probability of 0.75). We
conclude that HARP effectively overcomes the three error profiling challenges
introduced by on-die ECC.
\end{abstract}



\maketitle

\setlength{\footskip}{35pt} 

\fancyhf{}
\fancypagestyle{firstpage}
{
  \begin{tikzpicture}[remember picture,overlay]
    \node [xshift=142mm,yshift=-12mm]
    at (current page.north west) {\href{https://www.acm.org/publications/policies/artifact-review-and-badging-current}{\includegraphics[width=2.2cm]{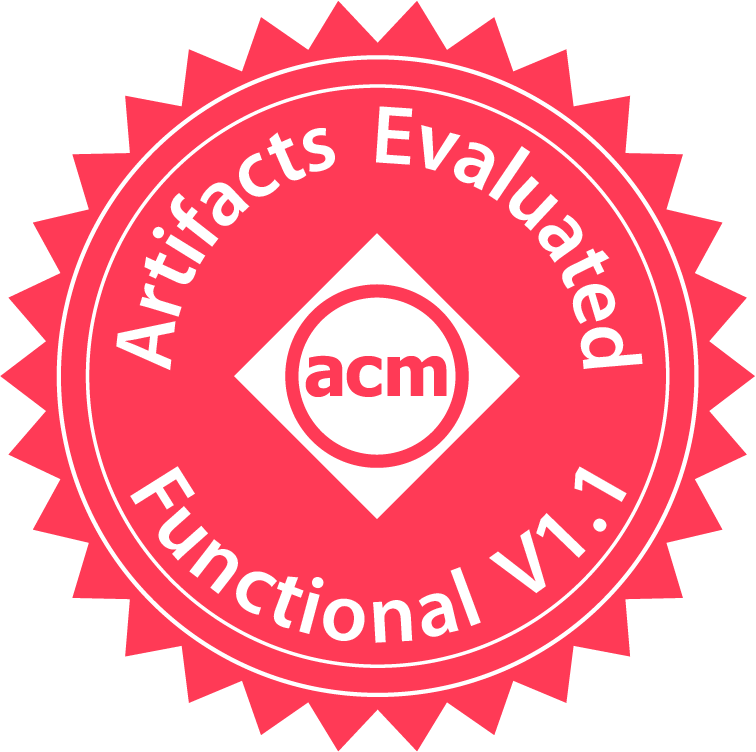}}} ;
    \node [xshift=165mm,yshift=-12mm]
    at (current page.north west) {\href{https://www.acm.org/publications/policies/artifact-review-and-badging-current}{\includegraphics[width=2.2cm]{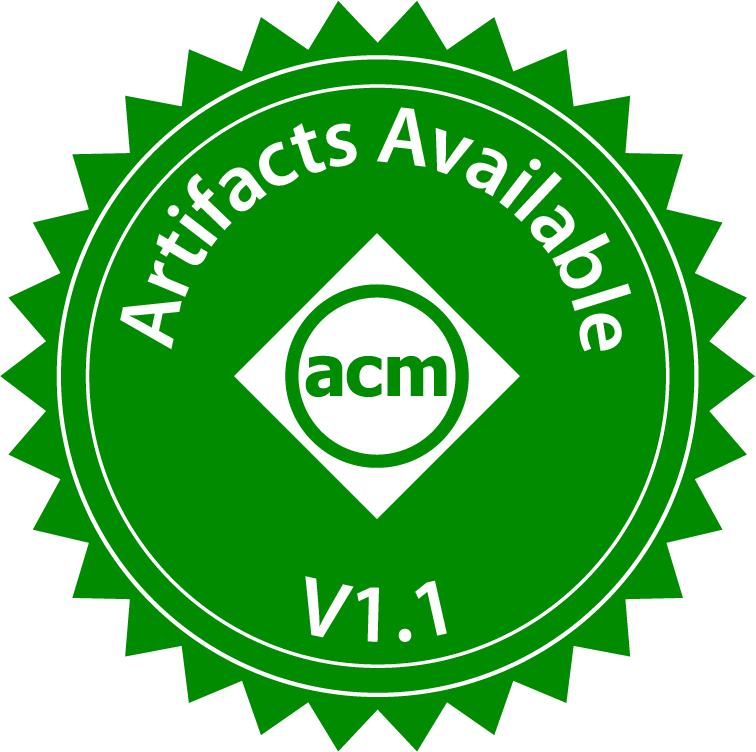}}} ;
    \node [xshift=188mm,yshift=-12mm]
    at (current page.north west) {\href{https://www.acm.org/publications/policies/artifact-review-and-badging-current}{\includegraphics[width=2.2cm]{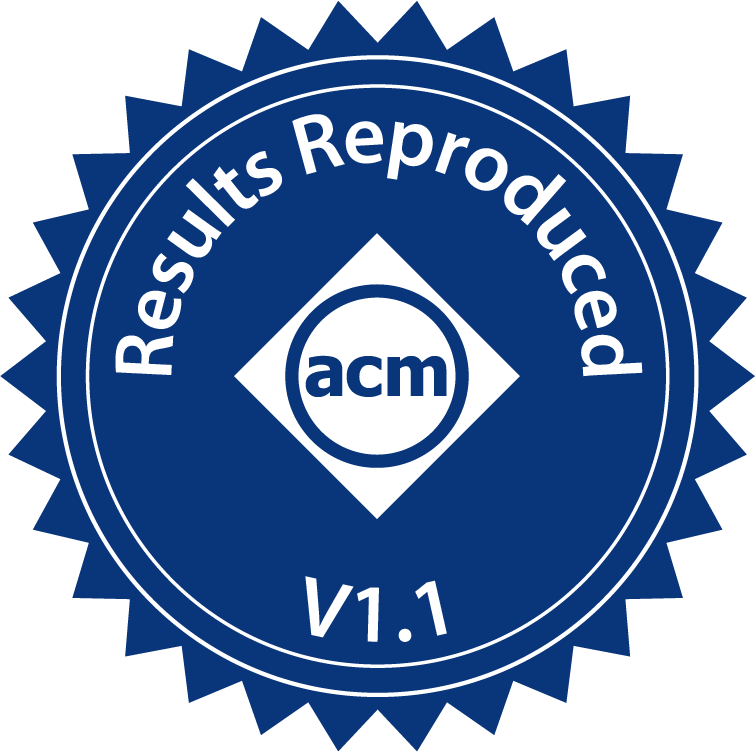}}} ;
\end{tikzpicture}

  \renewcommand{\headrulewidth}{0pt}
  \pagenumbering{arabic}
  \fancyfoot[C]{\large\thepage}
}

\fancypagestyle{otherpagestyle}
{
  \renewcommand{\headrulewidth}{0pt}
  \pagenumbering{arabic}
  \fancyfoot[C]{\large\thepage}
}

\thispagestyle{firstpage}
\pagestyle{otherpagestyle}

\ifcameraready
\else

  \def\parsepdfdatetime#1:#2#3#4#5#6#7#8#9{%
    \def\theyear{#2#3#4#5}%
    \def\themonth{#6#7}%
    \def\theday{#8#9}%
    \parsepdftime
  }

  \def\parsepdftime#1#2#3#4#5#6#7\endparsepdfdatetime{%
    \def\thehour{#1#2}%
    \def\theminute{#3#4}%
    \def\thesecond{#5#6}%
    \ifstrequal{#7}{Z}
    {%
      \def\thetimezonehour{+00}%
      \def\thetimezoneminute{00}%
    }%
    {%
      \parsepdftimezone#7%
    }%
  }

  \def\parsepdftimezone#1'#2'{%
    \def\thetimezonehour{#1}%
    \def\thetimezoneminute{#2}%
  }

  \newcommand*{\thetimezone}{\thetimezonehour:\thetimezoneminute}
  \expandafter\parsepdfdatetime\pdfcreationdate\endparsepdfdatetime

  \settimeformat{ampmtime}
  \newcommand{\versionnum}[0]{13.0}
  \newcommand{\version}[1]{\emph{Version #1 (Built:~\today~@ \currenttime~UTC\thetimezone)}}

  \AtBeginShipout
  {\AtBeginShipoutAddToBox{
      \begin{tikzpicture}[overlay, remember picture]
      \node[anchor=north] at (current page.north) {\textcolor{blue}{\vspace{3em}\version{\versionnum}}};    
      \end{tikzpicture}
  }}
\fi


\section{Introduction}
\label{sec:introduction}

Modern memory technologies that are suitable for main memory \mpm{(e.g., Dynamic
Random Access Memory (DRAM)~\cite{dennard1968field, mandelman2002challenges},
Phase-Change Memory (PCM)~\gfii{\cite{lee2009architecting, wong2010phase,
burr2010phase, lee2010phase,qureshi2009scalable}}, Spin-Transfer Torque RAM
(STT-RAM)~\cite{huai2008spin, kultursay2013evaluating})} all suffer from various
error mechanisms that play a key role in determining reliability, manufacturing
yield, and operating characteristics such as performance and energy
efficiency~\cite{lee2016technology, zhao2014improving, itrs2015more,
burr2010phase, wang2017architecting, lee2009architecting,
kultursay2013evaluating, kang2014co, park2015technology, cha2017defect,
kim2020revisiting, mandelman2002challenges, lee2010phase, guo2017sanitizer}.
Unfortunately, as memory designers shrink \mpm{(i.e., scale)} memory process
\mpm{technology node sizes} to meet \mpm{ambitious capacity, cost, performance,
and energy efficiency} targets, worsening reliability becomes an increasingly
significant challenge to surmount~\cite{lee2016technology, awasthi2012efficient,
mutlu2013memory, vatajelu2018state, cha2017defect, park2015technology,
kang2014co, nair2013archshield, qureshi2015avatar, kim2014flipping,
mutlu2019rowhammer, kim2020revisiting, hong2010memory, mandelman2002challenges,
kim1998dram}. For example, DRAM process technology \mpm{scaling} exacerbates
cell-to-cell variation and noise margins, severely impacting error mechanisms
that constrain yield, including cell data-retention~\cite{liu2013experimental,
nair2013archshield, kline2020flower, khan2016parbor, kang2014co,
venkatesan2006retention, hamamoto1995well, hamamoto1998retention,
park2015technology, cha2017defect, hong2010memory, patel2017reaper,
khan2014efficacy, shirley2014copula} and
read-disturb~\mpv{\cite{kim2014flipping, kim2020revisiting, mutlu2019rowhammer,
mutlu2017rowhammer, frigo2020trrespass, orosa2021deeper, hassan2021utrr}}
phenomena. Similarly, emerging main memory technologies suffer from various
error mechanisms that can lead to high error rates if left unchecked, such as
limited endurance, resistance drift, and write disturbance in
PCM~\cite{lee2009architecting, lee2009study, kim2005reliability, itrs2015more,
kang2006a, awasthi2012efficient} and data retention, endurance, and read
disturbance in STT-RAM~\cite{vatajelu2018state, chen2010advances,
apalkov2013spin, naeimi2013sttram, chun2012scaling, raychowdhury2009design}.
Therefore, enabling reliable system operation in the presence of scaling-related
memory errors is a critical research \mpq{challenge} for allowing continued main
memory scaling.

\textbf{Error mitigation \gfii{mechanisms} and on-die ECC.} Modern systems
\mpm{tolerate errors} using \emph{error-mitigation mechanisms}, which prevent
errors that occur within the memory \mpq{chip} from manifesting as
software-visible bit flips. Different error-mitigation mechanisms target
different types of errors, ranging from fine- to coarse-grained mitigation using
hardware and/or software techniques. \cref{subsec:scaling_errors} reviews main
memory \mpu{error-mitigation mechanisms.}

To address increasing memory error rates, memory chip manufacturers \mpq{have
started} to incorporate \emph{on-die error-correcting codes} (on-die
ECC).\footnote{Our work applies to \emph{any} memory \mpq{chip} that is packaged
with a proprietary ECC mechanism; on-die ECC is one embodiment of such a chip.}
On-die ECC is already prevalent among modern DRAM (e.g.,
LPDDR4~\cite{micron2017whitepaper, oh2014a, kwak2017a, kwon2017an,
patel2019understanding, patel2020bit}, DDR5~\cite{jedec2020ddr5}) and
STT-RAM~\cite{everspin2021sttmram} chips because it enables memory manufacturers
to aggressively scale their technologies while maintaining the appearance of a
reliable memory chip. Unfortunately, on-die ECC changes how memory errors appear
outside the memory chip (e.g., to the memory controller \mpq{or the system
software}). This introduces new challenges for \mpu{designing} additional
error-mitigation mechanisms at the system level~\mpv{\cite{son2015cidra,
gong2018duo, nair2016xed, jeong2020pair, cha2017defect, pae2021minimal,
criss2020improving, luo2014characterizing}} or test a memory chip's reliability
characteristics~\cite{patel2019understanding, patel2020bit, gold2014providing,
gorman2015memory}.

In this work, we focus on enabling a class of state-of-the-art hardware-based
error-mitigation mechanisms known as \emph{repair mechanisms}\footnote{We
discuss repair mechanisms in detail in
\cref{subsec:repair_mech}.}~\cite{yoon2011free, nair2013archshield,
ipek2010dynamically, schechter2010use, lin2012secret, tavana2017remap,
kline2017sustainable, longofono2021predicting, kline2020flower, nair2019sudoku,
seong2010safer, wilkerson2008trading, zhang2017dynamic, liu2012raidr,
qureshi2015avatar, venkatesan2006retention, son2015cidra} when used alongside
memory chips with on-die ECC. These repair mechanisms \mpm{operate from outside
the memory chip (e.g., from the memory controller)} to identify and repair
memory locations that are at risk of error (i.e., are known or predicted to
experience errors). In particular, prior works~\cite{nair2013archshield,
kline2017sustainable, tavana2017remap, schechter2010use, seong2010safer} show
that \emph{bit-granularity} repair mechanisms efficiently tackle high error
rates (e.g., $>10^{-4}$) resulting from aggressive technology scaling by
focusing error-mitigation resources on \mpq{bits} that are known to be
susceptible to errors.

\begin{figure}[H]
    \centering
    \includegraphics[width=\linewidth]{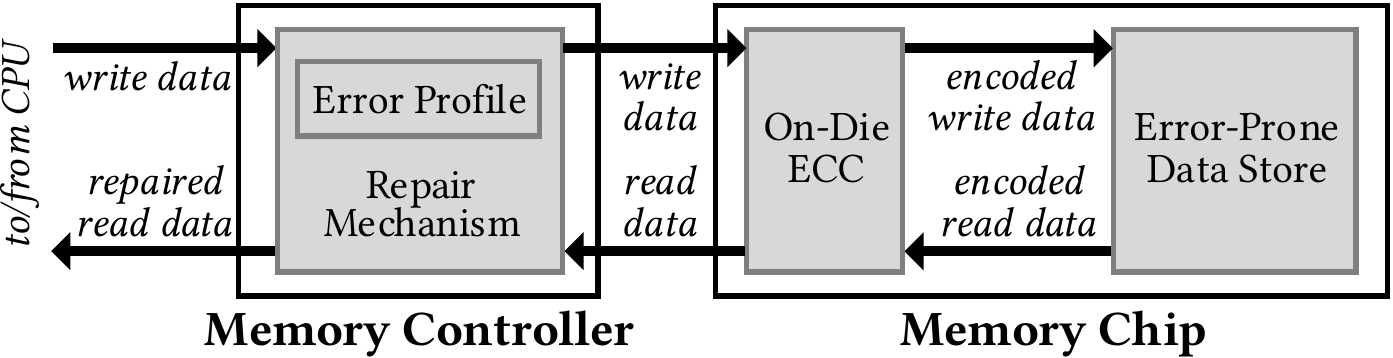}
    \caption{\mpm{High-level block diagram of a system that uses a repair mechanism with a memory chip that uses on-die ECC.}}
    \label{fig:sys_diag_simple}
  \end{figure}

\mpm{Fig.~\ref{fig:sys_diag_simple} illustrates a system that uses both a repair
mechanism (within the memory controller) and on-die ECC (within the memory
chip). \mpu{On-die ECC encodes all data provided by the CPU before writing it}
to the memory. On a read operation, \mpu{on-die ECC first decodes the stored
data, correcting any correctable errors. The repair mechanism then repairs the
data before returning it} to the CPU. The repair mechanism performs repair
operations \mpu{using} a list of bits known to be at risk of error, \mpu{called}
an \emph{error profile}.}

\textbf{Error profiling.}
\mpq{Repair} mechanisms depend on having a practical algorithm for identifying
at-risk memory locations to repair. We refer to this as an \emph{error profiling
algorithm}. \mpm{We classify a profiling algorithm as either} \emph{active} or
\emph{reactive} depending on whether it takes action to search for at-risk
memory locations or passively monitors memory to identify errors as they occur
during normal system operation. Prior works~\cite{venkatesan2006retention,
liu2012raidr, khan2014efficacy, khan2016case, khan2016parbor, khan2017detecting,
patel2017reaper, qureshi2015avatar, choi2020reducing, lee2017design,
bacchini2014characterization, sharifi2017online, kim2020revisiting,
frigo2020trrespass, cojocar19exploiting, zhang2012memory, hamdioui2017test,
tavana2017remap, qureshi2011pay, patel2020bit, liu2013experimental,
cojocar2020are, mutlu2019rowhammer, chang2017understanding, lee2015adaptive,
kim2018solar, kim2019d, baek2014refresh, kim2003block, lin2012secret} propose a
variety of algorithms for \emph{active} profiling. In general, these algorithms
all search for errors using multiple \emph{rounds} of tests that each attempt to
induce and identify errors (e.g., by taking exclusive control of the memory
chip~\cite{tavana2017remap, liu2013experimental, patel2017reaper,
patel2020bit}). \mpm{The algorithms maximize the chance of observing errors to
identify as many at-risk bits as possible by testing under \emph{worst-case
conditions} (e.g., data and access patterns, operating conditions)}. Only
BEEP~\cite{patel2020bit} accounts for the effects of on-die ECC by first
reverse-engineering the on-die ECC implementation, so we refer to all other
active profiling algorithms as \emph{Naive} in the context of this work. In
contrast, proposals for \emph{reactive} profiling passively monitor an
error-detection mechanism (typically an ECC) to identify errors as they occur
during normal system operation~\cite{jacob2010memory, mukherjee2004cache,
qureshi2015avatar, sharifi2017online, qureshi2011pay, meza2015large,
bajura2007models, saleh1990reliability}.

Regardless of the profiling algorithm, any at-risk bits that the profiler misses
will not be repaired by the repair mechanism that the profiler supports.
Therefore, achieving practical and effective repair requires a profiling
algorithm that \mpq{quickly and efficiently} achieves high coverage of at-risk
memory locations.

\textbf{On-die ECC's impact on error profiling.}
Unfortunately, on-die ECC \emph{fundamentally changes} how memory errors appear
outside of the memory chip: instead of \mpu{errors} that follow well-understood
semiconductor error models~\cite{jedec2016failure}, the system observes
\mpu{obfuscated error patterns that vary with the particular on-die ECC
implementation}~\cite{patel2019understanding, patel2020bit}. This is a serious
challenge for existing profiling algorithms because, as
\cref{sec:impl_on_error_prof} \mpu{shows}, on-die ECC both (1) \emph{increases}
the number of at-risk bits that need to be identified and (2) makes those bits
\mpu{harder} to identify. Even a profiler that knows the on-die ECC
implementation (e.g., BEEP~\cite{patel2020bit}) cannot easily identify all
at-risk bits because it \mpu{lacks} visibility into the error-correction
process.

\textbf{Our goal} is to study and address the challenges that on-die ECC
introduces for \mpm{bit-granularity error profiling}. To this end, we
\mpu{perform} the first analytical study of how \mpu{on-die ECC affects error
profiling}. We \mpu{find} that on-die ECC introduces statistical dependence
between errors in different bit positions such that, even if raw bit errors
(i.e., \emph{pre-correction} errors) occur independently, errors observed by the
system (i.e., \emph{post-correction} errors) do not. This raises three \mpm{new}
challenges for practical and effective bit-granularity error profiling
(discussed in detail in \cref{sec:impl_on_error_prof}).

First, on-die ECC transforms a small set of bits at risk of pre-correction
errors into a \emph{combinatorially larger} set of bits at risk of
post-correction errors. \cref{subsec:combinatorial_explosion} shows how this
exponentially increases the number of bits the profiler must identify. Second,
on-die ECC ties post-correction errors to specific combinations of
pre-correction errors. Only when those specific pre-correction error
combinations occur, can the profiler identify the corresponding bits at risk of
post-correction errors. \cref{subsec:bootstrapping} shows how this significantly
slows down profiling. Third, on-die ECC interferes with commonly-used memory
data patterns that profilers use to maximize the chance of observing errors.
This is because post-correction errors appear only when \emph{multiple}
pre-correction errors occur concurrently, which the data patterns must account
for. \cref{subsec:mutli_bit_test_pattern} \mpu{discusses the difficulty of
defining new data patterns for use with on-die ECC.}

To address these three challenges, we introduce Hybrid Active-Reactive Profiling
(HARP), a new bit-granularity error profiling algorithm that operates within the
memory controller to support a repair mechanism \mpq{for} memory chips with
on-die ECC. HARP is based on two key insights. First, given that on-die ECC uses
systematic encoding (i.e., data bits are preserved one-to-one during ECC
encoding\footnote{Nonsystematic designs require additional decoding effort
(i.e., more logic operations) because data \emph{cannot} be read \emph{directly}
from stored values~\cite{zhang2015vlsichapter41}. This increases the energy
consumption of read operations and either reduces the overall read timing
margins available for other memory operations or increases the memory read
latency.}), there are only two possible types of post-correction errors: (1)
\emph{direct} errors, corresponding to pre-correction errors within the
systematically encoded data bits; and (2) \emph{indirect} errors, resulting from
\emph{mistaken} correction operations (i.e., \emph{miscorrections}) on-die ECC
performs for \emph{uncorrectable} errors. Second, because on-die ECC corrects a
fixed number $N$ of errors, \emph{at most} $N$ indirect errors can occur
\mpq{concurrently} (e.g., $N=1$ for a Hamming code~\cite{hamming1950error}).

Based on these insights, the key idea of HARP is to \mpr{reduce the problem of
profiling a chip \emph{with} on-die ECC into that of a chip \emph{without}
on-die ECC} by separately identifying direct and indirect errors. HARP consists
of two phases. First, an active profiling phase that \mpr{uses existing
profiling techniques to identify} bits at risk of direct errors \mpr{with the
help of} a simple modification to the on-die ECC read operation that allows the
memory controller to read raw data (but not \mpq{the on-die ECC metadata})
values. Second, a reactive profiling phase that safely identifies bits at risk
of indirect errors using a secondary $N$-error-correcting ECC within the memory
controller. The secondary ECC is used \emph{only} for reactive profiling:
\mpr{upon identifying an error}, the corresponding bit is marked as at-risk,
which signals the repair mechanism to repair the bit thereafter.

Prior work~\cite{patel2020bit} shows that knowing the on-die ECC's internal
implementation (i.e., its parity-check matrix)\footnote{Potentially provided
through manufacturer support, datasheet information, or previously-proposed
reverse engineering techniques~\cite{patel2020bit}.} enables \emph{calculating}
which post-correction errors a given set \mpu{of pre-correction errors can
cause.} Therefore, we introduce two HARP variants: HARP-Aware (HARP-A) and
HARP-Unaware (HARP-U), which do and do not know the parity-check matrix,
respectively. HARP-A does \emph{not} improve coverage \mpu{over} HARP-U because
it has no additional visibility into the pre-correction errors. However, HARP-A
demonstrates that, although knowing the parity-check matrix alone does
\emph{not} overcome the three profiling challenges, it does provide a head-start
for reactive profiling based on the results of active profiling.

We evaluate HARP in simulation relative to two state-of-the-art baseline error
profiling algorithms: \emph{Naive} (which represents the vast majority of
previous-proposed profiling algorithms~\cite{venkatesan2006retention,
liu2012raidr, khan2014efficacy, khan2016case, khan2016parbor, khan2017detecting,
patel2017reaper, qureshi2015avatar, choi2020reducing, lee2017design,
bacchini2014characterization, sharifi2017online, kim2020revisiting,
frigo2020trrespass, cojocar19exploiting, zhang2012memory, hamdioui2017test,
tavana2017remap, qureshi2011pay, patel2020bit, liu2013experimental,
cojocar2020are, mutlu2019rowhammer, chang2017understanding, lee2015adaptive,
kim2018solar, kim2019d, baek2014refresh, kim2003block, lin2012secret}) and
\emph{BEEP}~\cite{patel2020bit}. We show that HARP quickly achieves coverage of
\emph{all bits at risk of direct errors} while Naive and BEEP \mpu{are either
slower or unable} to achieve full coverage. For example, when there are 2/3/4/5
bits at risk of pre-correction error that each fail with probability 0.5,
HARP\footnote{HARP-U and HARP-A have identical coverage of bits at risk of
direct error.} achieves 99th-percentile\footnote{We report 99th percentile
coverage to compare against baseline configurations that do not achieve full
coverage within the maximum simulated number of profiling rounds (due to
simulation time constraints, as discussed in \cref{subsubsec:sim_strat}).}
coverage in 20.6\%/36.4\%/52.9\%/62.1\% of the profiling rounds required by the
best baseline algorithm. Based on our evaluations, we conclude that HARP
effectively overcomes the three profiling challenges. We publicly release our
simulation tools as open-source software on Zenodo~\cite{harp-artifacts} and
Github~\cite{harpgithub}.

\mpq{To demonstrate the end-to-end importance of having an effective profiling
mechanism, we also} perform a case study of how HARP, Naive, and BEEP profiling
can impact the overall \mpi{bit error rate} of a system equipped with \mpm{an
ideal bit-repair mechanism that perfectly repairs all identified at-risk bits}.
We show that, because HARP achieves full coverage of bits at risk of direct
errors, it \mpq{enables} the bit-repair mechanism to repair all errors.
\mpq{Although Naive eventually achieves full coverage, it takes substantially
longer to do so (by $3.7\times$ for a raw per-bit error probability of 0.75). In
contrast, BEEP does \emph{not} achieve full coverage,} so the bit-repair
mechanism is unable to repair all errors that occur \mpq{during system
operation}.

We make the following contributions:
\begin{itemize}
\item We conduct the first analytical study of how \mpu{on-die ECC affects
system-level bit-granularity error profiling.} We identify three key challenges
that must be addressed to enable practical and effective \mpu{error profiling}
in main memory chips with on-die ECC.

\item We introduce Hybrid Active-Reactive Profiling (HARP), a new
bit-granularity error profiling \mpr{algorithm} that quickly achieves full
coverage of at-risk bits by profiling \mpq{errors} in two phases: (1) active
profiling using raw data bit values to efficiently identify all bits at risk of
direct errors; and (2) reactive profiling with the guarantee that all remaining
at-risk bits can be safely identified.

\item We introduce two HARP variants, HARP-Unaware (HARP-U) and HARP-Aware
(HARP-A). \mpm{HARP-A} exploits knowledge of the on-die ECC parity-check matrix
to achieve full coverage of at-risk bits \mpu{faster by precomputing at-risk bit
locations}.

\item We evaluate \mpm{HARP} relative to two baseline profiling algorithms.
\mpm{An example result shows} that HARP achieves 99th-percentile coverage of all
\mpm{bits at risk of direct error} in 20.6\%/36.4\%/52.9\%/62.1\% of the
profiling rounds required by the best baseline \mpm{technique} given 2/3/4/5
pre-correction errors.

\item We present a case study of how HARP and the two baseline profilers impact
the overall bit error rate of a system equipped with an \mpm{ideal repair
mechanism that perfectly repairs all identified at-risk bits}. We show that HARP
enables the repair mechanism to mitigate all errors \mpq{faster than the
best-performing baseline (e.g., by $3.7\times$ for a raw per-bit error
probability of 0.75)}.

\end{itemize}
\section{Background and Motivation}
\label{sec:background}

\mpu{This section briefly discusses} repair mechanisms, error profiling, our
assumed error model, and on-die ECC. For more detailed background information,
we refer the reader to other publications on
repair~\cite{horiguchi2011nanoscale, yoon2011free, nair2013archshield,
ipek2010dynamically, schechter2010use, lin2012secret, tavana2017remap,
kline2017sustainable, longofono2021predicting, kline2020flower, seong2010safer,
wilkerson2008trading}, profiling~\cite{venkatesan2006retention, liu2012raidr,
khan2014efficacy, khan2016case, khan2016parbor, khan2017detecting,
patel2017reaper, qureshi2015avatar, choi2020reducing, lee2017design,
bacchini2014characterization, sharifi2017online, kim2020revisiting,
frigo2020trrespass, cojocar19exploiting, zhang2012memory, hamdioui2017test} and
main memory error mechanisms~\cite{liu2013experimental, nair2013archshield,
kline2020flower, khan2016parbor, yaney1987meta, hamamoto1995well,
hamamoto1998retention, kim2014flipping, kim2020revisiting, lee2009architecting,
lee2009study, kim2005reliability, itrs2015more, kang2006a, vatajelu2018state,
chen2010advances, apalkov2013spin, naeimi2013sttram, chun2012scaling,
raychowdhury2009design, meza2015revisiting}.

\subsection{\rev{Addressing Scaling-Related Errors}}
\label{subsec:scaling_errors}

Continual increases to memory storage density exacerbate various
technology-specific error mechanisms (e.g., DRAM data
retention~\mpp{\cite{liu2013experimental, nair2013archshield, kline2020flower,
khan2016parbor, kang2014co, venkatesan2006retention, hamamoto1995well,
hamamoto1998retention, park2015technology, cha2017defect, hong2010memory,
patel2017reaper, khan2014efficacy, shirley2014copula, son2015cidra})} that
result in increasing error rates. \mpj{To address these errors,} main memory
manufacturers have already begun to \mpr{use} on-die ECC (e.g.,
LPDDR4~\mpp{\cite{oh2014a, kwon2017an, kwak2017a,
micron2017whitepaper}}, DDR5~\cite{jedec2020ddr5},
STT-RAM~\cite{everspin2021sttmram}) as a black-box error mitigation technique
\mpr{within the memory chip}, regardless of whether \mpp{or not} it is the most
efficient solution for a given system. On-die ECC \mpj{addresses} uncorrelated
single-bit errors that limit a manufacturers' factory
yield~\mpp{\cite{gong2018duo, micron2017whitepaper, im2020im, son2015cidra,
kang2014co, cha2017defect, patel2020bit, patel2019understanding}} and is already
prevalent among commodity DRAM \mpq{chips} today. Therefore, it is imperative
that system-level error-mitigation \mpp{mechanisms} take on-die ECC into
account, as \mpp{clearly} motivated by several prior
works~\mpp{\cite{son2015cidra, gong2018duo, nair2016xed, jeong2020pair,
cha2017defect, patel2020bit, criss2020improving}}.

\subsubsection{\rev{Mitigating High Error Rates.}}
\label{subsec:mitigating_high_bers}

As memory technologies continue to scale, \mpj{prior works~\cite{lin2012secret,
nair2013archshield, kline2017sustainable, nair2019sudoku, cha2017defect,
son2015cidra} argue that \mpp{raw bit} error rates will grow to \mpp{very} high
values (e.g., $>10^{-4}$ in DRAM~\cite{nair2013archshield,
cha2017defect}).}\footnote{\rev{Exact error rate values \mpq{of real memory
chips} are proprietary secrets because they \mpp{can reveal manufacturing
details and/or information relating to market
competitiveness}~\cite{gong2017dram, nair2013archshield}. Prior works draw
reasonable estimates from circuit metrics, e.g., coefficient of
variation~\cite{ipek2010dynamically, kline2020flower, nair2019sudoku}.}}
Enabling \mpp{robust} memory operation at these error rates is an established
research area for future DRAM and nonvolatile memories~\cite{lin2012secret,
nair2013archshield, kline2020flower, longofono2021predicting,
kline2017sustainable, schechter2010use, nair2019sudoku, zhang2017dynamic,
wang2017architecting, son2015cidra, david2011memory, mutlu2014research} because
it allows for both continued memory density scaling and enables new system
operating points and use-cases (e.g., reliably reducing memory timings and
voltages~\mpp{\cite{chang2017understanding, chang2016understanding,
chandrasekar2014exploiting, jung2016approximate, koppula2019eden,
zhang2016restore, naji2015high, shin2015dram, lee2017design, kim2018solar}})
that are not feasible today. \mpu{In general,} fine-grained (e.g., bit or word
granularity) hardware repair mechanisms can mitigate \mpu{high error rates} more
efficiently than conventional hardware error-mitigation techniques (e.g.,
ECC)~\cite{kline2017sustainable,longofono2021predicting, kline2020flower,
nair2013archshield, lin2012secret, qureshi2015avatar, son2015cidra}.
\mpj{\cref{subsec:repair_mech} discusses repair mechanisms in further detail.}

\subsubsection{\mpq{Consolidating In-Memory and Memory Controller Error Mitigations.}}

Unlike Flash memory~\gfii{\cite{cai2017error, cai2012error, luo2018improving, luo2018heatwatch, cai2018errors}}, main
memory is generally designed separately from the \mpp{memory
controller~\cite{mutlu2015main}. \mpq{Unfortunately, this separation discourages
building a unified error-mitigation mechanism across the memory and its
controller. This is exemplified by the widespread use of proprietary DRAM on-die
ECC, which introduces new reliability} challenges for designing error mitigation
mechanisms within the DRAM controller~\cite{gong2018duo, nair2016xed,
patel2020bit, cha2017defect, son2015cidra, criss2020improving}.} In general, the
standardized interface between the \mpq{memory and the controller} (e.g., JEDEC
DRAM standards~\cite{jedec2021main, jedec2020ddr5, jedec2012ddr4}) must be
modified to develop \mpq{a joint solution}, which impacts all manufacturers and
consumers involved, \mpq{and thus is a laborious and long (and often
politically-charged) process.} Therefore, we and other state-of-the-art
hardware-based repair mechanisms focus on minimizing the changes required to the
interface or memory chips themselves~\cite{nair2013archshield,
qureshi2015avatar,lin2012secret, kim2015balancing, kim2016relaxfault,
son2015cidra}.

\subsubsection{\mpj{Synergy with Other Error Mitigation \mpp{Approaches}.}}

Effective main memory error management is a \mpp{large} research space with
solutions spanning the entire hardware-software stack. There are many promising
solution directions, including software-driven repair techniques such as
\mpp{page retirement~\cite{mcelog2021bad, baek2014refresh, nvidia2020dynamic,
venkatesan2006retention, meza2015revisiting, hwang2012cosmic}} and
software-assisted techniques such as post-package repair
(PPR)~\cite{jedec2012ddr4, jedec2020ddr5, kang2014co, mandelman2002challenges,
nair2013archshield, son2015cidra, horiguchi2011nanoscale, cha2017defect}.
Unfortunately, these mechanisms have limitations that make them ill-suited to
address the high error rates that we target. For example, page \mpp{retirement}
operates at a coarse (i.e., system memory page) granularity, so it both wastes
significant capacity to repair the many errors at high error rates and cannot
easily repair in-use pages~\cite{meza2015revisiting, lee2019exploiting,
mcelog2021bad}. PPR provides only a few spare rows (e.g., one per bank in
DDR4~\cite{micron2020tn, kim2016relaxfault}) and suffers \mpp{from similar
drawbacks as page retirement due to operating at a coarse granularity
\mpq{(i.e., DRAM row)}}. In contrast, hardware-based repair mechanisms represent
the state-of-the-art in addressing scaling-related main memory errors.

\mpv{We note that other} error mitigation techniques may be used synergistically
with \mpq{hardware-based} repair, potentially to address different error models
\mpv{simultaneously}. \mpj{Our work both (1) identifies the challenges that
on-die ECC introduces for repair mechanisms; and (2) demonstrates a concrete way
(i.e., HARP) to address these challenges} with minimal changes to existing
systems. Given that on-die ECC is highly prevalent today (e.g.,
LPDDR4~\cite{oh2014a, micron2017whitepaper}, DDR5~\cite{jedec2020ddr5},
STT-RAM~\cite{everspin2021sttmram}), our ideas are applicable to and evaluated
based on current state-of-the-art solutions. In doing so, we believe our work
will help guide future solutions that develop new abstractions to step beyond
\mpq{simple} on-die ECC as it exists today.

\subsection{\rev{Enabling Repair Alongside On-Die ECC}}
\label{subsec:repair_mech}

Repair mechanisms~\cite{horiguchi2011nanoscale, yoon2011free,
nair2013archshield, ipek2010dynamically, schechter2010use, lin2012secret,
tavana2017remap, kline2017sustainable, longofono2021predicting, kline2020flower,
seong2010safer, wilkerson2008trading, mandelman2002challenges, son2015cidra,
kim2018solar, lee2017design} perform repair at granularities ranging from
kilobytes to single bits. The granularity at which \mpr{a repair mechanism
identifies} at-risk locations \mpr{is its} \emph{profiling granularity}. For
example, on-die row and column sparing~\cite{jedec2012ddr4, jedec2020ddr5,
kang2014co, mandelman2002challenges, nair2013archshield, son2015cidra,
horiguchi2011nanoscale, cha2017defect} requires identifying at-risk locations at
(or finer than) the granularity of a single memory row.
Table~\ref{tab:categorization} categorizes key repair mechanisms based their
profiling granularities. In general, coarse-grained repair requires less
intrusive changes to the system datapath because repair operations can align
with data blocks in the datapath (e.g., DRAM rows, cache lines, processor
words). However, this means that the repair mechanism suffers from more internal
fragmentation because each repair operation sacrifices more memory capacity
regardless of how \mpq{few} bits are actually at risk of error.

\begin{table}[H]
  \centering
  \scriptsize
 \setlength{\tabcolsep}{2pt}
  \begin{tabular}{llL{5cm}}
   \textbf{Profiling Granularity} & \textbf{Size (Bits)} & \textbf{Examples} \\\hline
          System page & 32 K & RAPID~\cite{venkatesan2006retention}, RIO~\cite{baek2014refresh}, Page~retirement~\cite{mcelog2021bad, nvidia2020dynamic, venkatesan2006retention, baek2014refresh, hwang2012cosmic} \\\hline
          \mpp{DRAM external row} & 2-64 K & PPR~\cite{jedec2012ddr4, jedec2020ddr5, kang2014co, mandelman2002challenges,
          nair2013archshield, son2015cidra, horiguchi2011nanoscale, cha2017defect}, Agnos~\cite{qureshi2015avatar}, RAIDR~\cite{liu2012raidr}, DIVA~\cite{lee2017design} \\\hline
          \mpp{DRAM internal row/col} & 512-1024 & Row/col sparing~\cite{kang2014co,
          mandelman2002challenges, nair2013archshield, son2015cidra,
          horiguchi2011nanoscale, cha2017defect}, Solar~\cite{kim2018solar} \\\hline
          Cache block  & 256-512  & FREE-p~\cite{yoon2011free}, CiDRA~\cite{son2015cidra} \\\hline
          Processor word  & 32-64  & ArchShield~\cite{nair2013archshield} \\\hline
          Byte & 8 & DRM~\cite{ipek2010dynamically} \\\hline
          Single bit & 1 &  ECP\footnotemark~\cite{schechter2010use},
          SECRET~\cite{lin2012secret}, REMAP~\cite{tavana2017remap},
          SFaultMap~\cite{kline2017sustainable}, HOTH~\cite{longofono2021predicting},
          FLOWER~\cite{kline2020flower}, SAFER~\cite{seong2010safer}, Bit-fix~\cite{wilkerson2008trading} \\\hline
  \end{tabular}
  \caption{Survey of \mpq{prevalent} memory repair mechanisms.}
  \label{tab:categorization}
\vspace{-2.5\baselineskip}
\end{table}
\footnotetext{\rev{ECP corrects individual bits, but its pointer size can be
adjusted to different granularities as required.}}

\begin{figure}[H]
    \centering
    \includegraphics[width=0.85\linewidth]{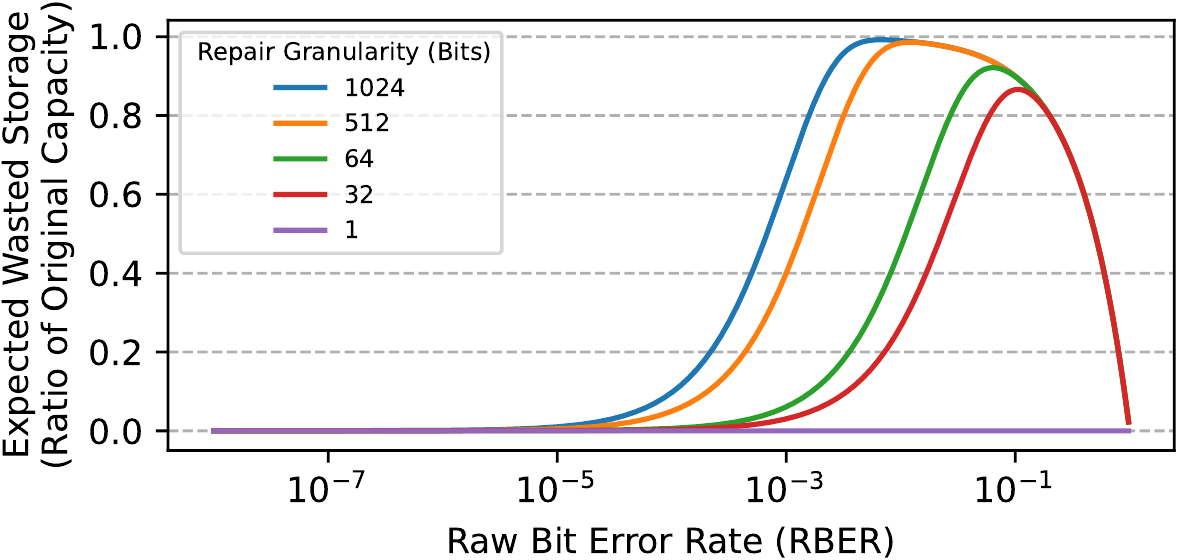}
    \caption{Expected amount of \mpq{wasted storage capacity} when \mpq{repairing} single-bit errors at various \mpq{repair} granularities.}
    \label{fig:wasted_cap}
    \vspace{-1em}
\end{figure}

Because of this tradeoff between intrusiveness and fragmentation, finer repair
granularities \mpj{are more efficient at higher error
rates~\cite{nair2013archshield, son2015cidra, cha2017defect}}.
Fig.~\ref{fig:wasted_cap} \mpj{illustrates this by showing} the expected
proportion of unnecessarily repaired bits (i.e., the amount of non-erroneous
memory capacity that is sacrificed alongside truly erroneous bits due to
internal fragmentation) (y-axis) at various raw bit error rates (x-axis) when
mitigating uniform-random single-bit errors at different repair granularities.
We see that coarse-grained repair becomes extremely wasteful as errors become
more frequent, e.g., wasting over 99\% of total memory capacity in the worst
case for a 1024-bit granularity \mpq{at a raw bit error rate of $6.8\times
10^{-3}$}. Note that the expected wasted storage decreases once the error rate
is sufficiently high because an increasing proportion of bits become truly
erroneous, \mpq{which reduces the wasted bits} for each repair operation. In
contrast, bit-granularity repair \mpq{(denoted with the line for `1')} does
\emph{not} suffer from internal fragmentation. For this reason, repair
mechanisms \mpq{designed for} higher error rates generally employ
finer-granularity profiling and repair~\cite{lin2012secret,
kline2017sustainable, longofono2021predicting, kline2020flower}.

\subsection{Practical and Effective Error Profiling}

\mpj{Any repair mechanism's effectiveness strongly depends on the effectiveness
of the error profiling algorithm that it uses because the repair mechanism can
only repair memory locations that it knows are at risk of error. In this
section, we define the key properties of practical and effective active and
reactive profilers.}

\subsubsection{\mpj{Active Profiling.}}

Active profiling algorithms take exclusive control of a memory \mpp{chip} in
order to (possibly destructively) test worst-case data and access
patterns~\cite{venkatesan2006retention, liu2012raidr, khan2016case,
khan2016parbor, khan2017detecting, patel2017reaper, lee2017design,
bacchini2014characterization, kim2020revisiting, frigo2020trrespass,
cojocar19exploiting, zhang2012memory, hamdioui2017test}, \mpu{so} the system
cannot perform useful work \mpu{while profiling}. Therefore, an active profiler
\mpu{must} identify at-risk bits as quickly and comprehensively as possible. We
quantify this by measuring the \mpu{fraction} of all at-risk bits that a given
profiler \mpu{identifies (i.e., its \emph{coverage})} across a fixed number of
testing rounds.

\subsubsection{\mpj{Reactive Profiling.}}

Reactive profiling algorithms (e.g., ECC scrubbing~\cite{han2014data,
qureshi2015avatar, choi2020reducing, sharifi2017online, alameldeen2011energy,
naeimi2013sttram}) \mpp{passively monitor error-detection mechanisms during
normal system operation, so their performance and energy impact is relatively
low and can be amortized across runtime~\cite{han2014data, qureshi2015avatar}.}
However, because \mpu{reactive profilers operate during \mpv{runtime}, they}
must ensure that they can not only detect, but also correct any errors that
occur. \mpp{Any errors that are not detected and corrected by the reactive
profiler risk introducing failures to the rest of the system.} In our work, we
quantify \mpq{the error-mitigation capability of a reactive profiler in terms of
ECC correction capability}, which is well-defined for ECCs used in memory
hardware design (i.e., linear block codes~\cite{roth2006introduction,
richardson2008modern, horiguchi2011nanoscale}).

\subsection{Errors and Error Models}
\label{subsec:errors_and_models}

Our work assumes uncorrelated single-bit errors because recent experimental
studies and repair efforts from academia~\cite{nair2013archshield,
bautista2016unprotected, longofono2021predicting, nair2019sudoku, son2015cidra},
industry~\cite{kang2014co}, and memory manufacturers
themselves~\cite{micron2017whitepaper, im2020im, ipek2010dynamically,
cha2017defect, park2015technology} focus on single-bit errors as the primary
reliability challenge with increasing storage density. In particular, DRAM and
STT-RAM manufacturers \mpr{use} on-die ECC specifically to combat these errors
in recent high-density chips~\cite{gong2018duo, micron2017whitepaper, im2020im,
kang2014co, oh2014a, everspin2021sttmram, cha2017defect}. Therefore, we assume
that errors exhibit the following properties:

\begin{enumerate}
    \item \emph{Bernoulli process:} independent of previous errors.
    \item \emph{Isolated:} independent of errors in other bits.
    \item \emph{Data-dependent:} dependent on the stored data pattern.
\end{enumerate}

To first order, this \mpj{error model suits} a broad range of error mechanisms
that relate to technology scaling and motivate the use of bit-granularity
repair, including DRAM data-retention~\cite{venkatesan2006retention,
hamamoto1995well, hamamoto1998retention, liu2013experimental,
nair2013archshield, kline2017sustainable, kline2020flower, khan2016parbor,
baek2014refresh, patel2017reaper, weis2015retention, jung2017platform,
jin2005prediction, kim2009new, kong2008analysis, lieneweg1998assesment} and read
disturbance~\cite{kim2014flipping, kim2020revisiting, park2016statistical}; PCM
endurance, resistance drift, and write disturbance~\cite{schechter2010use,
lee2009architecting, lee2009study, kim2005reliability, itrs2015more, kang2006a};
and STT-RAM data retention, endurance, and read
disturbance~\cite{vatajelu2018state, chen2010advances, apalkov2013spin,
naeimi2013sttram, chun2012scaling, raychowdhury2009design}. 
We use DRAM data-retention errors in our evaluations as a well-studied and
relevant example. However, a profiler is fundamentally agnostic to the
underlying error mechanism; \mpq{it identifies} at-risk bits based on whether or
not they \mpj{are observed to fail during} profiling. Therefore, our analysis
applies directly to any error mechanism that can be described using \mpq{the
aforementioned three properties.}

\noindent
\textbf{Correlated Errors.} Prior DRAM studies show evidence of correlated
errors~\mpv{\cite{al2005dram, meza2015revisiting, synopsys2015whitepaper,
sridharan2012study, sridharan2013feng, sridharan2015memory}}. However, we are
not aware of evidence that such errors are a first-order concern of modern DRAM
technology scaling. Correlated errors often result from faults outside the
memory array~\cite{meza2015revisiting} and are mitigated using fault-specific
error-mitigation mechanisms (e.g., write CRC~\cite{kwon2014understanding,
jedec2012ddr4}, chipkill~\cite{dell1997white, amd2009bkdg, nair2016xed} or even
stronger rank-level ECC~\cite{synopsys2015whitepaper, jedec2012ddr4,
kim2015bamboo}).

\noindent
\textbf{Low-Probability Errors.} Other main memory error mechanisms exist that
do not conform to our model, including time-dependent errors such as DRAM
variable retention time~\cite{yaney1987meta, restle1992dram, mori2005origin,
shirley2014copula, liu2013experimental, khan2014efficacy, qureshi2015avatar} and
single-event upsets such as particle strikes~\cite{may1979alpha}. In general,
these errors are either (1) inappropriate to address using a profile-guided
repair mechanism, e.g., single-event upsets that do not repeat; or (2) rare or
unpredictable enough that no realistic amount of \mpp{active} profiling is
likely to identify them, so they are left to \mpp{reactive profiling for
detection and/or mitigation}~\cite{qureshi2015avatar}. Identifying
low-probability errors is a general challenge for \emph{any} \mpq{error}
profiler and is an orthogonal problem to our work. Prior approaches to
identifying low-probability errors during \mpp{active} (e.g., increasing the
probability of error~\cite{patel2017reaper}) or \mpp{reactive (e.g., periodic
ECC scrubbing~\cite{awasthi2012efficient, qureshi2015avatar}) profiling} are
complementary to our proposed techniques and can be combined with HARP (e.g.,
during the active profiling phase \mpp{described in
\cref{subsec:active_prof_phase}, or by strengthening the secondary ECC as
described in \cref{subsubsection:incomplete_coverage}) to help identify
low-probability errors.}

\subsection{Block Codes and Syndrome Decoding}
\label{subsec:block_codes}

Typical on-die ECC implementations use linear block
codes~\cite{micron2017whitepaper, oh2014a, kwak2017a, kwon2017an, im2016im,
son2015cidra, cha2017defect, jeong2020pair} whose operation can be summarized
using matrix arithmetic. In our work, we assume single-error correcting (SEC)
Hamming codes~\cite{hamming1950error} that are adopted in modern
LPDDR4~\cite{oh2014a, micron2017whitepaper} and
DDR5~\cite{jedec2020ddr5} DRAM chips.\footnote{Our analysis can theoretically
generalize to stronger block codes (e.g., \mpp{double-error correcting}
BCH~\cite{bose1960class, hocquenghem1959codes}), but we leave \mpq{such
generalization} to future work given that such codes are \mpq{currently}
unlikely to be used \mpq{in} latency-sensitive memory chips.} This
section briefly summarizes the operation of an SEC Hamming code in the context
of on-die ECC. For further information, we refer the reader to extensive
literature on error-correction coding~\cite{richardson2008modern,
roth2006introduction, moon2005error, costello1982error}.

\subsubsection{Encoding and Decoding.}
\label{subsubsec:ecc_encoding_and_decoding}

\begin{figure}[hb]
  \centering
  \includegraphics[width=\linewidth]{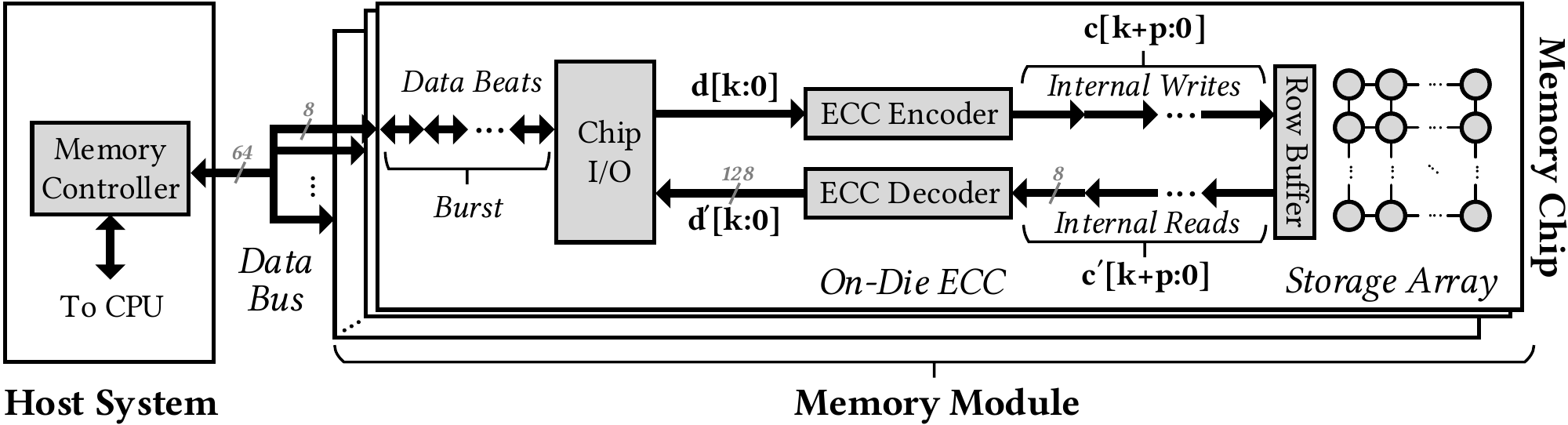}
  \caption{\rev{On-die ECC operation within a memory chip. Grey annotations show
  example bit-widths of data transfers.}}
  \label{fig:system_diagram}
  \vspace{-1em}
\end{figure}

A Hamming code with $k$ data bits and $p$ parity-check bits is defined by a $(k,
k+p)$ generator matrix $G$ and a $(p, k+p)$ parity-check matrix $H$ such that $G
\cdot H^T=\textbf{0}$ within the finite field $GF(2)$. Equation~\ref{eqn:hg}
gives example $H$ and $G$ matrices that define a $k=4$ SEC Hamming code.
\begin{equation}
\arraycolsep=0.9pt
\def\arraystretch{0.5}
\small
\mathbf{G^T}=\left[
  \begin{array}{cccc|ccc}
    1 & 0 & 0 & 0~ & ~1 & 1 & 1 \\
    0 & 1 & 0 & 0~ & ~1 & 1 & 0 \\
    0 & 0 & 1 & 0~ & ~1 & 0 & 1 \\
    0 & 0 & 0 & 1~ & ~0 & 1 & 1
  \end{array}
\right]
\quad\quad
\mathbf{H}=\left[
  \begin{array}{cccc|ccc}
    1 & 1 & 1 & 0~ & ~1 & 0 & 0 \\
    1 & 1 & 0 & 1~ & ~0 & 1 & 0 \\
    1 & 0 & 1 & 1~ & ~0 & 0 & 1
  \end{array}
\right]
\label{eqn:hg}
\end{equation}
Fig.~\ref{fig:system_diagram} illustrates how a system interfaces with a
memory \rev{module whose chip(s)} use on-die ECC. To encode a $k$-bit
\emph{dataword} $d$, the \emph{ECC encoder} computes a $(k+p)$-bit
\emph{codeword} $c$ as $c = G \cdot d$. Upon incurring raw bit error(s), we
denote the erroneous codeword as $c'$. To decode $c'$ into a post-correction
dataword $d'$, the \emph{ECC decoder} performs \emph{syndrome decoding}, where a
\emph{syndrome} $s$ is calculated as $s = H \cdot c'$. If $s$ is nonzero, one or
more errors must be present, and the bit position of the detected error can then
be identified as the particular column of $H$ that matches $s$. 

If an uncorrectable error occurs, $s$ may inadvertently match a
\mpp{parity-check matrix column} that does \emph{not} correspond to an actual
error. \mpp{In this case, the ECC decoder might introduce an \emph{additional}
error in the decoded data, which we refer to as a \emph{miscorrection}.} Note
that the memory controller may interface with one or more memory chips at a
time, potentially spreading a single data block across multiple on-die ECC
words. We discuss this further in \cref{subsubsec:matching_ondie_and_secondary_ecc}.

\subsubsection{Design Degrees of Freedom.}
\label{susbubsec:design_deg_freedom}

When choosing $H$ for a particular implementation, the designer is free to
arrange its columns at will. Recent work~\cite{pae2021minimal} shows that some
column arrangements can lead to more miscorrections than others. However,
designers may choose column arrangements based on circuit latency, energy, or
area concerns, regardless of \mpp{each choice's effect on reliability}. In our
work, we assume that the code uses \emph{systematic encoding}, which requires
that $H$ and $G$ do \mpq{not} modify data bits during encoding (note the
identity submatrices in Equation~\ref{eqn:hg}) \mpp{but does not otherwise
constrain the column arrangement}. This encoding greatly simplifies the hardware
encoding and decoding circuitry and is a realistic assumption for low-latency
main memory chips~\cite{zhang2015vlsichapter41}.
\section{\mpu{Formalizing Error Profiling}}
\label{sec:motivation2}
\label{subsubsec:expressing_error_models}

We express error profiling as a statistical process to understand \mpq{the
effects of} on-die ECC. To this end, we first formalize \mpq{the concepts of}
errors and error profiling. Then, we \mpq{examine how on-die ECC changes the way
that errors appear outside of the memory chip.}

\subsection{Representing the Probability of Error}
\label{subsec:rep_prob_error}

We model memory as a one-dimensional bit-addressable array with address space
$\mathcal{A}$. To describe \mpu{errors} within this address space, we define two
Boolean random variables $D_i$ and $E_i$ that represent the data stored in bit
$i \in \mathcal{A}$ and whether or not the bit experiences an error,
respectively. Based on our discussion in \cref{subsec:errors_and_models}, we
model $E_i$ as a Bernoulli random variable that is independent of $E_{j\neq i}$
but dependent on the data $D_i$. In general, $E_i$ can depend on the data stored
in other cells $D_{j \neq i}$, which expresses how \mpu{a bit's} probability of
error \mpu{changes with the data stored in nearby cells}.
Equation~\ref{eqn:bernoulli} shows the resulting probability mass function,
parameterized by $p$, the probability that the bit will experience an error.
\begin{equation}
  P(E_i=x|D_i, D_j, \cdots) =
  \begin{cases}
    p(D_i, D_j, \cdots)  & \text{if $x=1$} \\
    1 - p(D_i, D_j, \cdots) & \text{if $x = 0$}
  \end{cases}
  \label{eqn:bernoulli}
\end{equation}
In general, \mpk{each bit has its own value of $p$ depending on its intrinsic
error characteristics. For example, prior work~\cite{patel2017reaper}
experimentally demonstrates that $p$ values are normally distributed across
different bits for DRAM data-retention errors}, i.e., $p\sim N(\mu,
{\sigma}^2)$, with \mpk{some} normal distribution parameters $\{\mu, \sigma\}$
\mpk{that depend on the particular} memory chip and operating conditions such as
temperature.

\subsection{Incorporating On-Die ECC}
\label{subsec:incorporating_ecc_in_prof}

With on-die ECC, we adjust our address space representation to include both
\emph{logical} bit addresses $\mathcal{A}$ as observed by the memory controller
and \emph{physical} bit addresses $\mathcal{B}$ within the memory storage array.
In general, $|\mathcal{B}|>|\mathcal{A}|$ because $\mathcal{B}$ includes
addresses for parity-check bits that are \emph{not} visible outside of the
memory chip. Next, we introduce two additional \mpq{Boolean random variables:
$D_a$ \mpp{(for \underline{d}ataword)} and $C_b$ \mpp{(for
\underline{c}odeword)} that refer to the data values of logical bit
$a\in\mathcal{A}$ and physical bit $b\in\mathcal{B}$ (i.e., before and after ECC
encoding)}, respectively. Boolean variables $E_a$ and $R_b$ represent whether
logical bit $a$ and physical bit $b$ experience post- and pre-correction errors,
respectively. Note that $E$ and $D$ represent the same information from
\cref{subsec:rep_prob_error}. We use $C'$ and $D'$ to refer to codeword
and dataword values, respectively, that may contain errors. 

On-die ECC determines $D'$ from $C'$ through \mpu{syndrome decoding} (described
in \cref{subsec:block_codes}) \mpu{using} the ECC parity-check matrix $H$
\gfii{comprised of} columns $H[k+p:0]$. The error syndrome is computed as
$s=H[k+p:0] \cdot R[k+p:0]$ (referred to as $H \cdot R$ to simplify notation).
Then, if $s$ matches the $i$'th column $H[i]$, the ECC decoder flips the bit at
position $i$. Given that $H$ is systematically encoded (discussed in
\cref{subsubsec:ecc_encoding_and_decoding}), $c[k:0]$ is equal to $d[k:0]$.
Therefore, a post-correction error $E_i$ (i.e., a mismatch between $d_i$ and
$d'_{i}$ ) can only occur in two cases: (1) an uncorrected raw bit error at
position $i$ (i.e., $R_i \wedge s\neq H[i]$); or (2) a miscorrection at position
$i$ (i.e., $\neg R_i \wedge s=H[i]$.) We refer to these two cases as
\emph{direct} and \emph{indirect} errors, respectively.
Equation~\ref{eqn:dr_error_post_ecc} summarizes both cases that lead to a
post-correction error.
\begin{align}
    P(E_i) &= P(R_i \veebar H \cdot R=H[i])
  \label{eqn:dr_error_post_ecc}
\end{align}
Equation~\ref{eqn:dr_error_post_ecc} \mpu{shows that bit $i$'s probability of
error depends \emph{not only} on that of its encoded counterpart $R_i$, but
\emph{also} on those of} \emph{all other codeword bits} $R[k+p:0]$. This means
that on-die ECC introduces statistical dependence between \emph{all bits} in a
given ECC word through their mutual dependence on $R$. \mpq{Furthermore, just as
described in \cref{subsec:rep_prob_error}, $R_i$ itself depends on the data
value stored in cell $i$ (i.e., $C_i$). As a result, \mpu{bit $i$'s probability
of error} depends on \emph{both} the data values and the pre-correction errors
\mpu{present} throughout the codeword.}

\mpq{Consequently, a given post-correction error $E_i$ may \emph{only} occur
when a particular combination of pre-correction errors occurs. This is different
from} the case without on-die ECC, where $E_i$ does \emph{not} depend on the
data or error state of any other bit \mpq{$j \neq i$}. We conclude that on-die
ECC transforms statistically independent pre-correction errors into
ECC-dependent, correlated post-correction errors.
\section{\mpu{On-Die ECC's Impact on Profiling}}
\label{sec:impl_on_error_prof}

On-die ECC breaks the simple and intuitive assumption that profiling for errors
is the same as profiling each bit individually. In this section, we identify
three key challenges that on-die ECC introduces for bit-granularity profiling.

\subsection{Challenge 1: Combinatorial Explosion}
\label{subsec:combinatorial_explosion}

\mpq{\cref{subsec:incorporating_ecc_in_prof} shows that the position of
an indirect error depends on the locations of all pre-correction errors $R$.
This means that different uncorrectable patterns of pre-correction errors can
cause indirect errors in different bits. In the worst case, \emph{every unique
combination} of pre-correction errors within a set of at-risk bits can lead to
different indirect errors. This means that the set of bits that are at risk of
post-correction errors is \emph{combinatorially larger} than the set of bits at
risk of pre-correction error.}

\begin{figure}[H]
  \centering
  \includegraphics[width=0.9\linewidth]{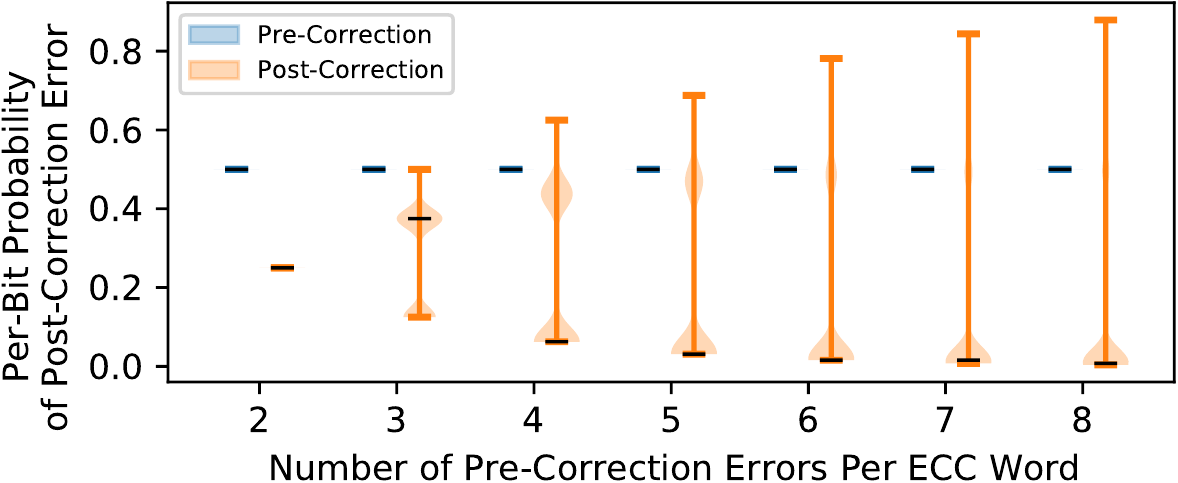}
  \caption{Distribution of \mpq{each at-risk bit's error probability} before and
  after application of on-die ECC.}
  \label{fig:pre_and_post_errors}
  \vspace{-1em}
\end{figure}

As a concrete example, Fig.~\ref{fig:pre_and_post_errors} shows a violin plot of
each at-risk bit's per-bit probability of error (y-axis) when the codeword
contains a fixed number of bits at risk of pre-correction errors (x-axis) that
each fail with probability $0.5$. Each violin shows the distribution
\mpu{(median marked in black)} of per-bit error probabilities when simulating
70,000 ECC words for each of 1600 randomly-generated (71, 64) Hamming code
parity-check matrices assuming a data pattern of 0xFF. We make two observations.
First, the pre-correction error probabilities are all 0.5 (by design). This
means that, without on-die ECC, all bits at risk of pre-correction error are
easy to identify, i.e., each bit will be identified with probability
$p=1-0.5^{N}$ given $N$ profiling rounds. For large $N$ (e.g., $N>10$, where $p>
0.999$), the vast majority of bits will be identified. 

\mpq{Second, in contrast, the per-bit probabilities of post-correction error
exhibit a wide range. However, the probability density for each violin is
tightly concentrated at $Y\approx 0.4$ for $X=3$ and shifts towards $Y=0$ as the
number of pre-correction errors increases. This means that the bits at risk of
post-correction errors become \emph{harder to identify} because they fail less
often.}

\mpq{Table~\ref{tab:amplification} shows the maximum number of bits at risk of
post-correction errors that can be caused by a fixed number of bits at risk of
pre-correction errors. This illustrates the worst-case scenario, where
\emph{every \mpr{uncorrectable} combination} of pre-correction errors (i.e.,
\emph{pre-correction error pattern}) causes a unique indirect error. We see that
$n$ bits at risk of pre-correction errors can cause $2^{n}-1$ unique
pre-correction error patterns. Of these, $n$ are correctable error patterns,
leaving $2^{n}-n-1$ \emph{uncorrectable pre-correction error patterns}. Assuming
that each of these patterns introduces a unique indirect error, the combination
of bits at risk of direct and indirect error leads to $2^n-1$ bits at risk of
post-correction errors. Therefore, we conclude that on-die ECC
\mpr{exponentially} increases the number of at-risk bits that the profiler must
identify.}

\begin{table}[h]
    \centering
    \small
    \setlength\tabcolsep{2.5pt}
    \def\arraystretch{1}
    \begin{tabular}{lllllll}
        \textbf{Bits at risk of pre-correction errors}
          & \boldmath{$n$} & \textbf{1} & \textbf{2} & \textbf{3} & \textbf{4} & \textbf{8}\\\hline
        Unique pre-correction error patterns
          & $2^n - 1$ & 1 & 3 & 7 & 15 & 255 \\
        Uncorrectable pre-correction error patterns
          & $2^n - n - 1$ & 0 & 2 & 4 & 11 & 247\\\hline
        \textbf{Bits at risk of post-correction errors}
          & \boldmath{$2^n - 1$} & \textbf{1} & \textbf{3} & \textbf{7} & \textbf{15} & \textbf{255}\\
    \end{tabular}
    \caption{On-die ECC amplifies a few bits at risk of pre-correction errors into exponentially many bits at risk of post-correction errors.}
    \label{tab:amplification}
    \vspace{-3em}
  \end{table}

\subsection{Challenge 2: Profiling without Feedback}
\label{subsec:bootstrapping}

\mpq{Without on-die ECC, an at-risk bit is identified when the bit fails. This
means} that every profiling round provides useful feedback about which bits
\emph{are} and \emph{are not} at risk of error. \mpq{Unfortunately, with on-die
ECC, a bit at risk of post-correction errors can \emph{only} be identified} when
\emph{particular combination(s)} of pre-correction errors occur. \mpq{This has
two negative consequences.}
 
\mpq{First, because the profiler cannot observe pre-correction errors, it} does
not know whether a particular combination of pre-correction errors has been
tested \mps{yet}. Therefore, the profiler \emph{cannot} draw meaningful
conclusions from observing a bit \emph{not} to fail. \mpq{Instead, the profiler
must pessimistically suspect every bit to be at risk of post-correction errors,
even after many \mps{profiling} rounds have elapsed without observing a given
bit fail.} \mps{Second, each ECC word can only exhibit \emph{one} pre-correction
error pattern at a time (i.e., during any given profiling round). This}
serializes the process of identifying any two bits at risk of post-correction
errors that fail under different pre-correction error patterns.

As a result, \mpu{no} profiler that identifies at-risk bits based on observing
post-correction errors \mpu{can} quickly identify all bits at risk of
post-correction errors. We refer to this problem as \emph{bootstrapping} because
the profiler must explore different pre-correction error patterns without
knowing which patterns it is exploring. In \cref{subsubsec:bootstrapping_anal},
we find that bootstrapping \gfii{limits the profiler's coverage of \mpu{at-risk
bits} to incremental improvements across profiling rounds.} 

\subsection{Challenge 3: Multi-Bit Data Patterns}
\label{subsec:mutli_bit_test_pattern}

Designing data patterns that induce worst-case circuit conditions is a difficult
problem that depends heavily on the particular circuit design of a given memory
chip and the error mechanisms it is susceptible to~\cite{dekker1990realistic,
cheng2002neighborhood, khan2018test, mrozek2019multi, mrozek2010analysis,
cui2016snake}. Without on-die ECC, a bit can fail \mpq{only} in one way, i.e.,
when it exhibits an error. Therefore, the worst-case pattern \mpq{needs to} only
consider factors that affect the bit itself (e.g., data values stored in the bit
and its neighbors).

Unfortunately, with on-die ECC, a given post-correction error can potentially
occur with multiple different pre-correction error patterns. Therefore, the
worst-case data pattern must both (1) account for different ways in which the
post-correction error can occur; and (2) for each way, consider the worst-case
conditions \mpq{for the individual pre-correction errors to occur
simultaneously}. This is a far more complex problem than without on-die
ECC~\cite{gold2014providing, gorman2015memory}, and in general, there may not
even be a single worst-case data pattern that exercises all possible cases in
which a given post-correction error might occur. To our knowledge, no prior work
has \mpq{developed a general solution to this problem,} and we identify this as
a key direction for future research.
\section{\mpu{Addressing the Three Challenges}}
\label{subsec:practical_prof}
\label{sec:addressing_challenges}

We observe that all three profiling challenges stem from the \emph{lack of
access} that the profiler has into pre-correction errors. Therefore, we conclude
that \emph{some} transparency into the on-die ECC mechanism is necessary to
enable practical error profiling in the presence of on-die ECC. \mpu{This
section discusses options for enabling} access to pre-correction errors and
\mpv{describes} our design choices for HARP.

\subsection{Necessary Amount of Transparency}

\mpn{To reduce the number of changes we require from the memory chip, we
consider the minimum amount of information that the profiler needs to make error
profiling \emph{as easy as} if there were no interference from on-die ECC. To
achieve this goal, we examine the following two insights that are derived in \cref{subsec:incorporating_ecc_in_prof}.}

\begin{enumerate}
    \item \mpn{Post-correction errors arise from either direct or indirect
    errors.}
    
    \item \mpn{The number of concurrent indirect errors is limited to the correction
    capability of on-die ECC.}
\end{enumerate} 

\mpn{First, we observe that it is \emph{not} necessary for the profiler to have
full transparency into the on-die ECC mechanism or pre-correction errors. If all
bits at risk of direct errors can be identified, all remaining indirect errors
are upper-bounded by on-die ECC's correction capability. Therefore, the indirect
errors can be safely identified from within the memory controller, e.g., using a
reactive profiler.}

\mpn{Second, we observe that the profiler can determine exactly which
pre-correction errors occurred within the data bits (though not the parity-check
bits) simply by knowing at which bit position(s) on-die ECC performed a
correction operation. This is because the data bits are systematically encoded
(\mpq{as} explained in \cref{susbubsec:design_deg_freedom}), so their
programmed values must match their encoded values. By observing which bits
experienced direct error(s), the profiler knows which pre-correction errors
occurred within the encoded data bits.}

Based on these observations, we require that the profiler be able to identify
which direct errors occur on every access, including \mpu{those that on-die ECC
corrects}. Equivalently, on-die ECC may expose the error-correction operation
that it performs so that the profiler can infer the direct \mpu{errors from the
post-correction data.}

\subsection{Exposing Direct Errors to the Profiler}

We consider two different ways to inform the profiler about pre-correction
errors within the data bits.

\begin{enumerate}

\item \emph{Syndrome on Correction.} On-die ECC reports the error syndrome
calculated on all error correction events, which corresponds to the bit
position(s) that on-die ECC corrects.

\item \emph{Decode Bypass.} On-die ECC provides a read access path that bypasses
error correction and returns the raw values stored in the data portion of the codeword. 
\end{enumerate}

\noindent 
We choose to build upon decode bypass because we believe it to be the easiest to
adopt for three key reasons. First, there exists precedent for similar on-die
ECC decode bypass paths from both academia~\cite{gong2018duo} and
industry~\cite{bains2020read} with trivial modifications to internal DRAM
hardware, and an on-die ECC disable configuration register is readily exposed in
certain DRAM datasheets~\cite{alliance2020lpddr4}. Second, prior works already
reverse-engineer both the on-die ECC algorithm~\cite{patel2020bit} and raw bit
error rate~\cite{patel2019understanding} without access to raw data bits or
insight into the on-die ECC mechanism, so we do not believe exposing a decode
bypass path reveals significantly more sensitive information. Third, we strongly
suspect that such a bypass path \emph{already exists} for post-manufacturing
testing~\cite{thun2020qualification}. This is reasonable because
systematically-encoded data bits can be read out directly without \mpv{requiring
further transformation.} If so, exposing this capability as a feature would
likely require minimal engineering effort for the potential gains of new
functionality. However, we recognize that the details of \mpn{the on-die ECC
implementation} depend on the particular memory chip design, and it is
ultimately up to the system designer to choose the most suitable option for
their system. 

\subsection{Applicability to Other Systems}

Any bit-granularity profiler operating without visibility into \mpn{the
pre-correction errors} suffers form the three profiling challenges \mpn{we
identify in this work}. Even a hypothetical future main memory system whose
memory chips and controllers are designed by the same (or two trusted) parties
will need to account for \mpq{and overcome} these profiling challenges when
incorporating a repair mechanism \mpq{that relies on practical and effective
profiling}.
\begin{figure*}[hb]
    \centering
    \includegraphics[width=0.95\textwidth]{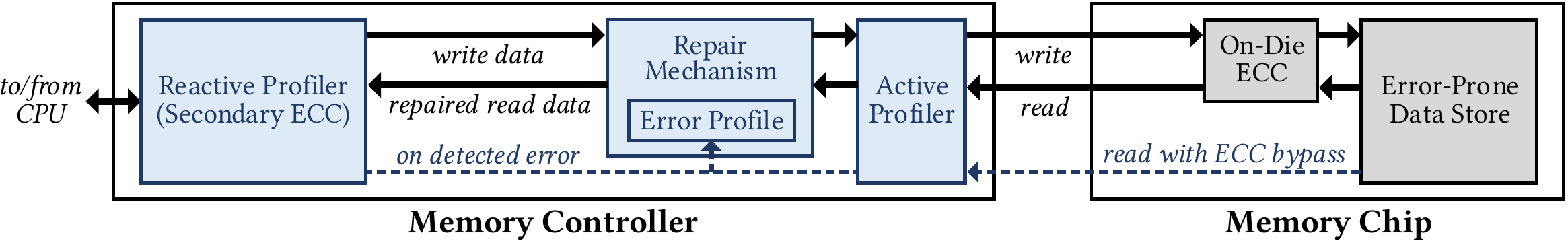}
    \caption{Block diagram summarizing the error-mitigation resources (in blue) of a HARP-enabled system.}
    \label{fig:harp_sys_diagram}
\end{figure*}

\section{\mpu{Hybrid Active-Reactive Profiling}}

We introduce the Hybrid Active-Reactive Profiling (HARP) algorithm, which
overcomes the three profiling challenges introduced by on-die ECC \mpq{discussed
in \cref{sec:addressing_challenges}. HARP separates profiling into \emph{active}
and \emph{reactive} phases that independently identify bits at risk of
\emph{direct} and \emph{indirect} errors, respectively.}

\subsection{HARP Design Overview}
\label{subsec:harp_design_overview}

\mpq{Fig.~\ref{fig:harp_sys_diagram} illustrates the high-level architecture
of a HARP-enabled system, with the required error-mitigation resources shown in
blue. The memory chip exposes a read operation with the ability to bypass on-die
ECC and return the raw data (though not parity-check) bits. The memory
controller contains a repair mechanism with an associated error profile
alongside both an \emph{active} and \emph{reactive} profiler. During active
profiling, the active profiler uses the ECC bypass path to search for bits at
risk of direct errors. Because the active profiler interfaces
\gfii{\emph{directly}} with the raw data bits, its profiling process is
equivalent to profiling a memory chip without on-die ECC. If and when direct
errors are observed, the active profiler communicates their locations to the
repair mechanism's error profile.}

\mpq{After active profiling is complete, the reactive profiler (i.e., a
secondary ECC code with correction capability \gfii{at least as strong as} that
of on-die ECC) continuously monitors for bits at risk of indirect errors. The
reactive profiler is responsible \emph{only} for identifying bits at risk of
indirect errors the first time that they fail. If and when the reactive profiler
identifies an indirect error, the location of the error is recorded to the error
profile for subsequent repair.}

\subsection{Active Profiling Implementation}
\label{subsec:active_prof_phase}

\mpq{The active profiler follows the general round-based algorithm employed by
state-of-the-art error profilers, as} discussed in
\cref{sec:introduction}. Each round of testing first programs memory
cells with a standard memory data pattern that is designed to maximize the
chance of observing errors (e.g., $0xFF$, $0x00$, random
data)~\gfii{\cite{adams2002high, mrozek2019multi, liu2013experimental, khan2016parbor, kim2020revisiting}}. Patterns may
or may not change across testing rounds depending on the requirements of the
particular data pattern. Once sufficiently many rounds are complete, the set of
at-risk bits identified comprises the union of all bits identified across all
testing rounds.

We assume that the active profiler achieves full coverage of bits at risk of
direct errors by leveraging any or all of the worst-case testing techniques
developed throughout prior works~\cite{patel2017reaper, khan2016parbor,
khan2017detecting, mukhanov2020dstress, mrozek2019multi}. \mpq{This is feasible
because the active profiler can read and write to the raw data bits
\mpr{exploiting} the ECC bypass path and the systematically-encoded data bits,
respectively. Therefore, the active profiler can use techniques developed for
memory chips without on-die ECC.}

\subsection{Reactive Profiling Implementation}
\label{subsubsec:matching_ondie_and_secondary_ecc}

HARP requires that the secondary ECC \mpq{have correction capability \emph{at
least} as high as the number of indirect errors that on-die ECC can cause at one
time. This requires \mpr{the} layout of secondary ECC words to account for the
layout of on-die ECC words: the two must combine in such a way that every on-die
ECC word is protected with the necessary correction capability \gfii{by the secondary ECC}. For example,
with a single-error correcting on-die ECC that uses 128-bit words, the memory
controller must ensure that every 128-bit on-die ECC word is protected with
\emph{at least} single-error correction.}

How this is achieved \mpu{depends heavily on a given system's memory
architecture}. For example, depending on the size of an on-die ECC word and how
many memory chips the memory controller interfaces with, on-die ECC words may be
split across different data transfers. In this case, the system designer must
choose a design that \mpu{matches} their design goals, e.g., \mpu{dividing}
secondary ECC words across multiple transfers (which introduces its own
reliability challenges~\cite{gong2018duo}), or interleaving secondary ECC words
across multiple on-die ECC words (which could require stronger secondary ECC).

\mpq{Without loss of generality, we assume that the memory controller interfaces
with a single memory chip at a time (e.g., similar to \mpr{some}
LPDDR4 systems~\gfii{\cite{jedec2014lpddr4}}) and provisions a single-error correcting
code per on-die ECC word. Such a system is sufficient for demonstrating the
error profiling challenges that we address in this work.} Matching the
granularity of secondary and on-die ECC words for \mpq{arbitrary systems is
\emph{not} a problem unique to our work since \emph{any} secondary ECC 
that \gfii{is designed} to account for the effects of on-die ECC must consider how the two
interact~\cite{cha2017defect, gong2018duo}. Therefore,} we leave a general
exploration of the tradeoffs \mpr{involved} to future work.

\subsubsection{HARP-U and HARP-A\gfii{.}}
\label{subsubsection:harp_u_and_harp_a}

\mpq{We introduce two variants of HARP: HARP-A and HARP-U, which are aware and
unaware of the on-die ECC parity-check matrix $H$, respectively. HARP-A uses
this knowledge to \emph{precompute}\footnote{Using the methods described in
detail in prior work~\cite{patel2020bit}.} bits at risk of indirect error given
the bits at risk of direct error that are identified during active profiling.
HARP-A does \mpq{not} provide benefits over HARP-U during active profiling.
However, HARP-A reduces the number of bits at risk of indirect error that remain
to be identified by reactive profiling.}

\subsubsection{Increasing the Secondary ECC Strength\gfii{.}}
\label{subsubsection:incomplete_coverage}

The secondary ECC is used for reactive profiling and must \mpq{provide equal or
greater correction capability} than on-die ECC to safely identify indirect
errors. Current on-die ECC designs are limited to simple single-error correcting
codes due to area, energy, and latency constraints within the memory
die~\cite{micron2017whitepaper, gong2018duo, son2015cidra}, so the secondary ECC
can be correspondingly simple. However, if either (1) future on-die ECC designs
become significantly more complex; or (2) \mpq{the system designer wishes to
address other failure modes (e.g., component-level failures) using the secondary
ECC, the system designer will need to increase the secondary ECC strength
accordingly}. Whether profile-based repair remains a feasible error-mitigation
strategy in this case is ultimately up to the system designer and their
reliability goals, so we leave further exploration to future work.

\subsection{Limitations}
\label{subsubsec:full_coverage}

HARP relies on the active profiler to achieve full coverage of bits at risk of
direct errors so that the reactive profiler \emph{never} observes a direct error
(i.e., the reactive profiler's correction capability is never exceeded).
Consequently, if the active profiler fails to achieve full coverage, the
reactive profiler \mpq{may experience indirect errors \emph{in addition to}
direct error(s) missed by active profiling.} 

\mpq{We acknowledge this as a theoretical limitation of HARP, but we do not
believe it restricts HARP's potential impact to future designs and scientific
studies. This is because achieving full coverage of at-risk bits without on-die
ECC is a long-standing problem that is complementary to our work. Prior works
have studied this problem in detail~\cite{khan2014efficacy, khan2017detecting,
qureshi2015avatar, patel2017reaper}, and any solution developed for chips
without on-die ECC can be immediately applied to HARP's active profiling phase,
effectively reducing the difficulty of profiling chips with on-die ECC to that
of chips without on-die ECC.}
\section{Evaluations}
\label{sec:evaluations}

In this section, we study how HARP's coverage of direct and indirect errors
changes with \mpn{different pre-correction error counts and per-bit error
probabilities} to both (1) demonstrate the effect of the three profiling
challenges introduced by on-die ECC and (2) show that HARP overcomes the three
challenges. 

\subsection{Evaluation Methodology}
\label{subsec:methodology}

We evaluate error coverage in simulation because, unlike with a real device, we
can accurately compute error coverage using precise knowledge of which errors
are and are not possible. This section describes our simulation methodology.

\subsubsection{Baselines for Comparison.}

We compare HARP-U and HARP-A with two baseline profiling algorithms that use
multiple rounds of testing \mpn{with different data patterns to identify at-risk
bits based only on observing post-correction errors.}

\begin{enumerate}
    \item \emph{Naive}, \mpr{which represents} the vast majority of prior
    profilers that operate without knowledge of on-die
    ECC~\cite{venkatesan2006retention, liu2012raidr, khan2014efficacy,
    khan2016case, khan2016parbor, khan2017detecting, patel2017reaper,
    qureshi2015avatar, choi2020reducing, lee2017design,
    bacchini2014characterization, sharifi2017online, kim2020revisiting,
    frigo2020trrespass, cojocar19exploiting, zhang2012memory, hamdioui2017test,
    tavana2017remap, qureshi2011pay, patel2020bit, liu2013experimental,
    cojocar2020are, mutlu2019rowhammer, chang2017understanding, lee2015adaptive,
    kim2018solar, kim2019d, baek2014refresh, kim2003block, lin2012secret}
    \mpr{(described in \cref{subsec:active_prof_phase})}.
    
    \item \emph{BEEP}, the profiling algorithm supported by the
    reverse-engineering methodology BEER~\cite{patel2020bit}. BEEP careful\gfi{ly}
    constructs data patterns intended to systematically expose post-correction
    errors based on having reverse-engineered the on-die ECC parity-check
    matrix. We follow the SAT-solver-based methodology as described by\gfi{~\cite{patel2020bit}} and use a random data pattern before the first post-correction error
    is confirmed.
\end{enumerate}
    
\subsubsection{Simulation Strategy.}
\label{subsubsec:sim_strat}

We extend the open-source DRAM on-die ECC analysis infrastructure released by
prior work~\gfi{\cite{beersimgithub,patel2020bit}} to perform Monte-Carlo
simulations of DRAM data retention errors. \mpn{We release our simulation tools
on Zenodo~\cite{harp-artifacts} and Github~\cite{harpgithub}.} We simulate error
injection and correction using single-error correcting Hamming
codes~\cite{hamming1950error} representative of those used in DRAM on-die ECC
\mpq{(i.e., (71, 64)~\cite{im2016im, nair2016xed} and (136, 128)~\cite{oh2014a, kwon2017an, kwak2017a, micron2017whitepaper} code configurations). All
presented data is shown for a (71, 64) code, and we verified that our
observations hold for longer (136, 128) codes.} We simulate 1,036,980 total ECC
words across 2769 randomly-generated parity-check matrices \mpq{over $\approx$20
days of simulation time (discussed in \cref{subsubsec:eval_runtime_estimate}). \mpq{For each profiler
configuration, we simulate 128 profiling rounds because this is enough to
understand the behavior of each configuration (e.g., the shapes of each curve in
Fig.~\ref{fig:eval_type1_coverage}), striking a good balance with simulation
time.}} 

We inject errors according to the model discussed in
\cref{subsec:errors_and_models} to simulate the effect of uniform-random,
data-dependent errors. We assume that all bits are
true-cells~\gfi{\cite{kraft2018improving, liu2013experimental}} that experience
errors \emph{only} when programmed with data `1', \mpq{which is consistent with
experimental observations made by prior work~\cite{kraft2018improving,
patel2020bit}}. To study how varying error rates impact profiling, we simulate
bit errors with Bernoulli probabilities of 1.0, 0.75, 0.5, and 0.25 and separate
our results based on the total number of pre-correction errors $n$ injected into
a given ECC word. Using this approach, one can \gfii{easily} determine the effect of
an arbitrary raw bit error rate by summing over the individual per-bit
\gfi{error} probabilities.

We define \emph{coverage} as the proportion of all at-risk bits that are
identified. We calculate coverage using the Z3 SAT solver~\cite{de2008z3},
computing the total number of post-correction errors that are possible for a
given (1) parity-check matrix; (2) set of pre-correction errors; and (3)
(possibly empty) set of already-discovered post-correction errors. Note that a
straightforward computation of coverage given on-die ECC is extremely difficult
for data-dependent errors: each data pattern programs the parity-check bits
differently, thereby provoking different pre-correction error patterns.
Therefore, using the SAT solver, we accurately measure the bit error rate
\mpr{of \emph{all possible} at-risk bits across} \emph{all possible} data
patterns.

We simulate three different \mpq{data} patterns to exercise data-dependent
behavior: \texttt{random}, \texttt{charged} \gfii{(i.e., all bits are `1's)}, and \texttt{checkered} \gfii{(i.e., consecutive bits alternate between `0' and `1')}. For the
\texttt{random} and \texttt{checkered} data patterns, we invert the data pattern
each round of profiling. For the \texttt{random} pattern, we change the random
pattern after every two profiling rounds (i.e., after both the pattern and its
inverse are tested). We ensure that each profiler is evaluated with the exact
same set of ECC words, pre-correction error patterns, and data patterns in order
to preserve fairness when comparing coverage values. Unless otherwise stated,
all data presented uses the \texttt{random} pattern, which we find performs on
par or better than the static \texttt{charged} and \texttt{checkered} patterns
that do \gfi{\emph{not}} explore different pre-correction error combinations.

\subsection{Active Phase Evaluation}

We study the number of profiling rounds required by each profiling algorithm to
achieve coverage of direct errors. We omit HARP-A because its coverage of direct
errors is equal to that of HARP-U.

\subsubsection{Direct Error Coverage.}
\label{subsubsec:direct_error_coverage}

\begin{figure}[H]
    \centering
    \includegraphics[width=\linewidth]{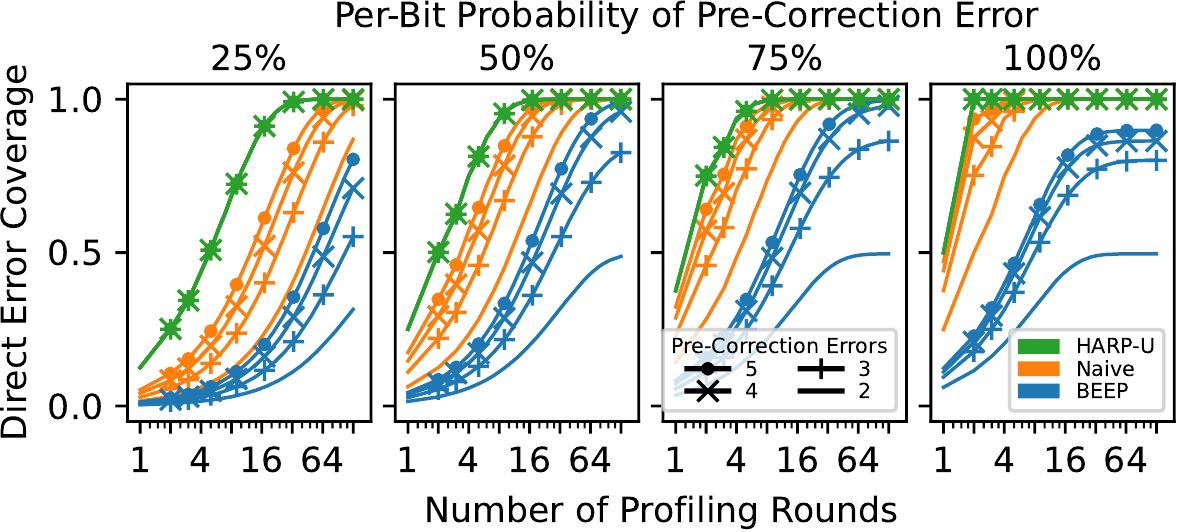}
    \vspace{-1em}
    \caption{Coverage of bits at risk of direct errors.}
    \label{fig:eval_type1_coverage}
    \vspace{-1em}
\end{figure}

Fig.~\ref{fig:eval_type1_coverage} shows the coverage of bits at risk of
direct errors that each profiler cumulatively achieves \gfi{(y-axis)} over 128
profiling rounds \gfi{(x-axis) assuming four different values of pre-correction
errors per ECC word (2, 3, 4, and 5). We report results for four different
per-bit error probabilities of the injected pre-correction errors (25\%, 50\%,
75\%, 100\%)}. For each data point, we compute coverage as the proportion of
at-risk bits identified out of all at-risk bits across all simulated ECC words. 

We make two observations. First, HARP consistently and quickly achieves full
coverage, \emph{regardless of} the number or per-bit error probabilities of the
injected pre-correction errors. This is because HARP bypasses on-die ECC
correction, identifying each at-risk bit independently, regardless of which
error occurs in which testing round. In contrast, both Naive and BEEP \gfii{(1)}
require more testing rounds to achieve coverage parity with HARP and \gfii{(2)}
exhibit significant dependence on the number of pre-correction errors. This is a
direct result of on-die ECC: each post-correction error depends on particular
combination(s) of pre-correction errors, and achieving high coverage requires
these combinations to occur in distinct testing rounds. We conclude that that
HARP effectively overcomes the first profiling challenge by directly observing
pre-correction errors, while Naive and BEEP must both rely on uncorrectable
error patterns to incrementally improve coverage \mpv{in} each round.

Second, although, HARP and Naive both \gfi{\emph{eventually}} achieve full
coverage, BEEP \gfi{\emph{fails}} to \mpu{do so} in certain cases. This is
because BEEP does not explore all pre-correction error combinations necessary to
expose each bit at risk of direct errors. We attribute this behavior to a nuance
of the BEEP algorithm: BEEP \mpu{crafts data patterns that increase the
likelihood of indirect errors. Unfortunately, these patterns are slow to explore
different combinations of pre-correction errors, which leads to incomplete
coverage}. This is consistent with \mpu{prior work~\cite{patel2020bit}, which
finds that BEEP exhibits low coverage when pre-correction errors are sparse or
occur with low probability. We find that Naive also fails to achieve full
coverage when using static data patterns (e.g., \texttt{checkered}) for the same
reason.}

\subsubsection{Bootstrapping Analysis.}
\label{subsubsec:bootstrapping_anal}

\begin{figure}[H]
    \centering
    \includegraphics[width=\linewidth]{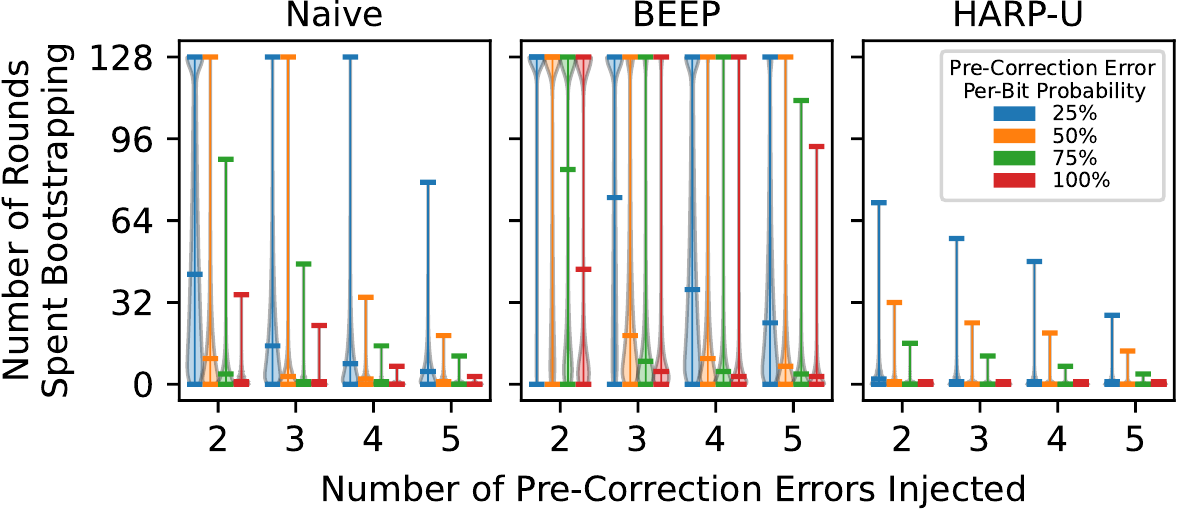}
    \caption{Distribution of the number of profiling rounds required to identify the first direct error across all simulated ECC words.}
    \label{fig:eval_bootstrapping}
\end{figure}

Fig.~\ref{fig:eval_bootstrapping} shows the distribution (median marked with a
horizontal line) of the number of profiling rounds required for each profiler
\mpu{to observe \emph{at least one} direct error in each ECC word. If} no
post-correction errors are identified, we conservatively plot the data point as
requiring 128 rounds, which is the maximum number of rounds evaluated (discussed
in \cref{subsubsec:sim_strat}). The data illustrates the difficulty of
bootstrapping because observing \mpv{any} post-correction error with on-die ECC
requires a \mpv{specific combination of} pre-correction errors to occur.

\gfi{We make three observations. First, w}e \gfi{see} that HARP identifies the
first error far more quickly than Naive or BEEP profiling across \emph{all}
configurations\gfi{. Second, HARP} never fails to identify at least one error
\gfi{within 128 rounds}. \gfi{Third, i}n contrast, BEEP sometimes \gfi{cannot}
identify an error at all due to a combination of (1) the low per-bit
pre-correction error probability and (2) the bootstrapping problem \gfi{(i.e.,}
more testing rounds does \emph{not} guarantee higher coverage unless those
rounds explore different uncorrectable patterns\gfi{)}. We conclude that HARP
effectively addresses the bootstrapping challenge by directly observing
pre-correction errors instead of relying on exploring different uncorrectable
error patterns.

\subsection{Reactive Phase Evaluation}

In this section, we study each \gfi{error} profiler's coverage of bits at risk
of indirect errors and examine the correction capability required from the
secondary ECC to safely identify the at-risk bits remaining after active
profiling. \mpq{\gfii{I}t is important to note that\gfii{, unlike HARP,} neither Naive nor BEEP
profiling achieve full coverage of bits at risk of \gfii{\emph{direct}} errors for all
configurations. In such cases, multi-bit errors can occur during reactive
profiling that are not safely identified by a single-error correcting code
(studied in \cref{subsubsec:missed_uncorr_events}), regardless of the
profiler's coverage of bits at risk of indirect errors.}

\subsubsection{Indirect Error Coverage.}

\begin{figure}[H]
    \centering
    \includegraphics[width=\linewidth]{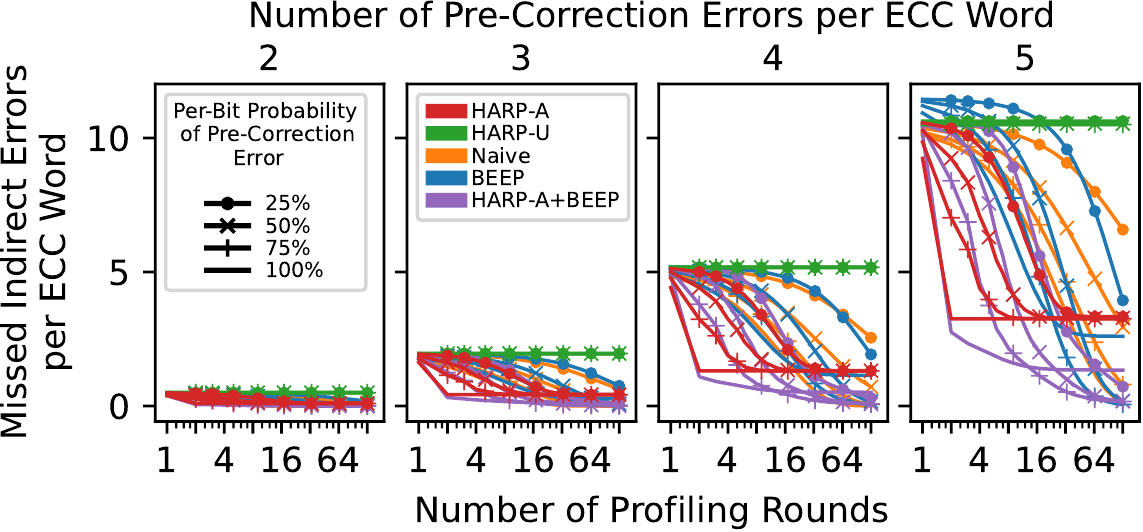}
    \caption{Coverage of bits at risk of indirect errors.}
    \label{fig:type_2_coverage}
\end{figure}

Fig.~\ref{fig:type_2_coverage} shows the proportion of all bits that are at
risk of indirect errors that each profiler \emph{has missed} per ECC word
throughout 128 rounds of profiling. This is equivalent to the number of at-risk
bits \gfi{that} reactive profiling \gfi{has} to identify. We evaluate an
additional configuration, HARP-A+BEEP, which employs BEEP to identify the
remaining at-risk bits once HARP-A has identified all bits at risk of direct
errors.

\begin{figure*}[!b]
    \centering
    \begin{subfigure}[b]{0.49\textwidth}
        \centering
        \includegraphics[width=\textwidth]{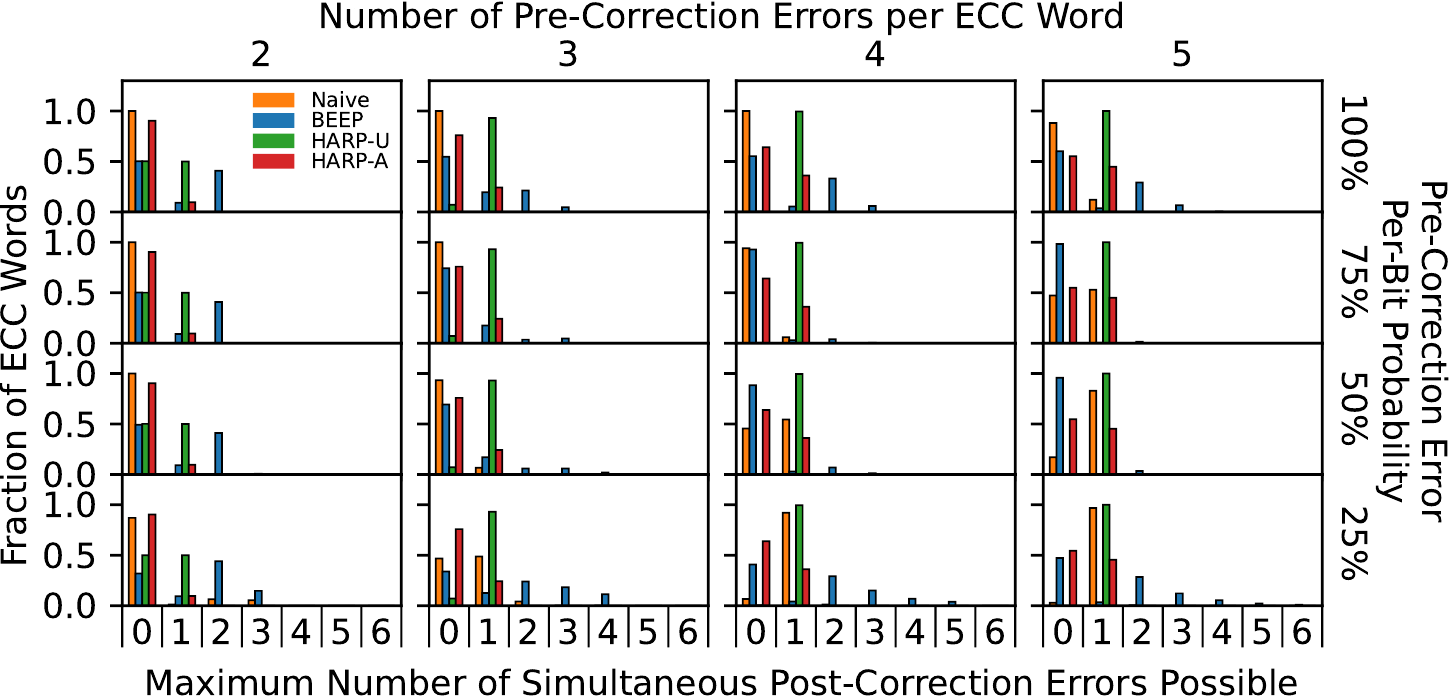}
        \caption{\mpq{Normalized histogram of the maximum number of simultaneous
        post-correction errors (x-axis) possible across all simulated ECC words
        after 128 rounds of profiling.}}
        \label{fig:max_simulteaneous_a}
    \end{subfigure}
    \hfill
    \begin{subfigure}[b]{0.49\textwidth}
        \centering
        \includegraphics[width=\textwidth]{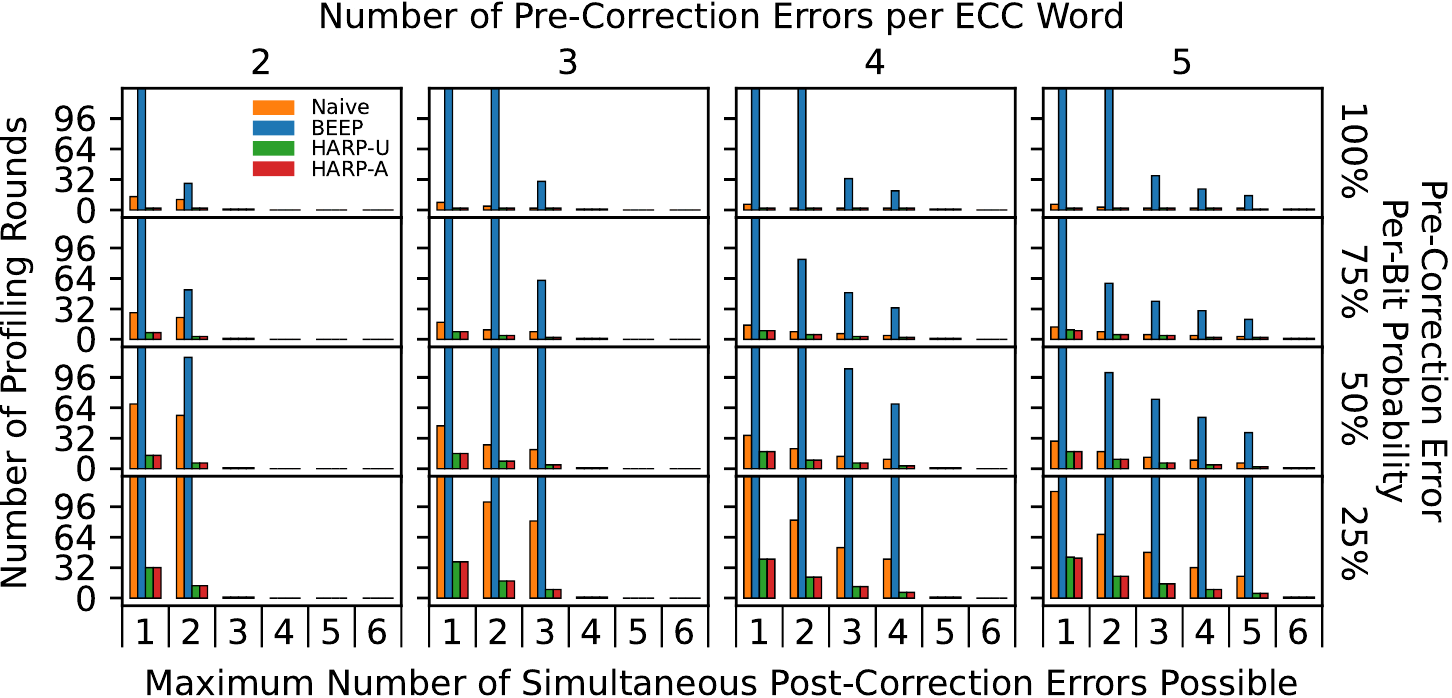}
        \caption{Number of profiling rounds (y-axis) required to achieve 99th-percentile values of the maximum number of simultaneous post-correction errors possible (x-axis).}
        \label{fig:max_simulteaneous_b}
    \end{subfigure}    
    \caption{Maximum number of simultaneous post-correction errors possible
    given all at-risk bits missed after 128 rounds of profiling.}
    \label{fig:max_simultaneous}
\end{figure*}

We make three observations. First, HARP-U does not identify any bits at risk of
indirect errors\footnote{Except for a small number of direct and indirect errors
that overlap.} because it \mpu{bypasses the} on-die ECC correction process that
causes indirect errors. In contrast, HARP-A quickly identifies a subset of all
bits at risk of indirect errors by predicting them from the \mps{identified}
direct errors. \mps{Note that HARP-A cannot} identify \emph{all} bits at risk of
indirect errors because \mps{doing so would require knowing which parity-check
bits are at risk of error, which the on-die ECC bypass path does not reveal.}

Second, combining HARP-A with BEEP effectively overcomes \mps{HARP-A's}
inability to identify pre-correction errors within the parity-check bits. This
is because HARP-A+BEEP synergistically combines \mpq{(1) HARP's ability to
quickly identify bits at risk of direct errors with (2) BEEP's ability to
exploit \emph{known} at-risk bits to \emph{expose} others.} The combined
configuration quickly identifies bits at risk of indirect errors, achieving
\mpq{coverage similar to Naive and BEEP profiling in less than half the number
of profiling rounds.}

Third, both Naive and BEEP achieve relatively high coverage of indirect errors
after many (i.e., $>64$) rounds compared to \mps{HARP-U and HARP-A}. This is
because \mps{both Naive and BEEP continually} explore different uncorrectable
error patterns, \mps{steadily exposing more and more indirect errors.} BEEP
achieves higher coverage because its \mpq{algorithm deliberately seeks} out
pre-correction error combinations that are more likely to cause post-correction
errors, thereby exposing more indirect errors \mps{in the long run}.

We conclude that knowing the on-die ECC parity-check matrix \mps{helps HARP-A
and BEEP identify} bits at risk of indirect errors, thereby reducing the number
of indirect errors that must be identified by \mpu{by the secondary ECC during}
reactive profiling.

\subsubsection{Secondary ECC Correction Capability.}
\label{subsubsec:missed_uncorr_events}

Fig.~\ref{fig:max_simultaneous} shows the worst-case (i.e., maximum) number of
post-correction errors that can occur simultaneously within an ECC word
\mpq{after active} profiling.
\mps{This number is} the correction capability required from \mpq{secondary ECC}
to safely \mpq{perform reactive profiling}.

\noindent
\textbf{Maximum Error Count.} Fig.~\ref{fig:max_simultaneous}a shows a
normalized histogram of the maximum number of post-correction errors that can
occur simultaneously within each simulated ECC word given all at-risk bits
missed after 128 rounds of active profiling. We observe that both HARP-U and
HARP-A exhibit \emph{at most one} post-correction error across all
configurations. This is because HARP \mpq{identifies all bits at risk of direct
errors within 128 profiling rounds (shown in
Fig.~\ref{fig:eval_type1_coverage}), so only one error may occur at a time
(i.e., an indirect error).} In contrast, both Naive and BEEP are susceptible to
multi-bit errors. In particular, BEEP's relatively low coverage of bits at risk
of direct errors means that many multi-bit error patterns remain possible.
\mpq{We conclude that\gfii{, after 128 rounds of active profiling,} a
single-error correcting secondary ECC is \gfii{\emph{sufficient}} to perform
reactive profiling for \gfii{HARP \mps{but} \gfii{\emph{insufficient} \mps{to do
so} for}} Naive and BEEP.}

\noindent
\textbf{Maximum Error Count.} Fig.~\ref{fig:max_simultaneous}b shows \mpq{how
many active profiling rounds are required to ensure that \emph{no more than} an
x-axis value of post-correction errors can occur simultaneously \mpu{in a single
ECC word} during reactive profiling. We conservatively report results for the
99th percentile of all simulated ECC words because neither Naive nor BEEP
achieve full coverage of bits at risk of direct errors for all configurations
within 128 profiling rounds.} In cases where 128 profiling rounds are
insufficient to achieve 99th-percentile values, we align the bar with the top of
the plot.

We make two observations. First, \mpu{both HARP configurations perform}
\emph{significantly} faster than Naive and BEEP. For example, with a 50\%
pre-correction per-bit error probability, HARP \mpq{ensures that no more than
one post correction error can occur in 20.6\%/36.4\%/52.9\%/62.1\% of the
profiling rounds required by Naive} given 2/3/4/5 pre-correction errors. This is
because HARP quickly identifies all bits at risk of direct errors, while Naive
and BEEP \mpv{both either (1) take longer to do so; or (2) fail to do so
altogether (e.g., for the 100th percentile at a 50\% per-bit error
probability)}. Second, BEEP performs much worse than any other profiler because
it exhibits extremely low coverage of bits at risk of direct error (studied in
\cref{subsubsec:direct_error_coverage}). We conclude that achieving high
coverage of bits at risk of direct errors is essential for minimizing the
correction capability of the secondary ECC.

\subsection{\rev{Case Study: DRAM Data Retention}}
\label{subsec:harp_case_study_data_retention}

\mpq{In this section, we show how \gfii{error} profiling impacts end-to-end reliability. We
study the bit error rate of a system that uses a bit-granularity repair
mechanism (e.g., such as those discussed in \cref{subsec:repair_mech}) to
reduce the DRAM refresh rate, which prior work shows can significantly improve
overall system performance and energy-efficiency~\cite{venkatesan2006retention,
liu2012raidr, qureshi2015avatar, patel2017reaper} and enable continued density
scaling~\cite{nair2013archshield}. We assume that data-retention errors follow
the error model described in \cref{subsec:errors_and_models} (i.e.,
uniformly with a fixed raw bit error rate, which is consistent with prior
experimental studies~\cite{sutar2016d, patel2017reaper, baek2014refresh,
hamamoto1998retention, kim2018dram, shirley2014copula, patel2019understanding})}
and that the repair mechanism perfectly repairs any at-risk bits that are
identified by either active or reactive profiling. We assume a (71, 64) SEC
on-die ECC code and a secondary ECC capable of detecting and correcting a single
error in each on-die ECC word during reactive profiling.

\begin{figure*}[t]
    \centering
    \begin{subfigure}[b]{0.49\textwidth}
        \centering
        \includegraphics[width=\textwidth]{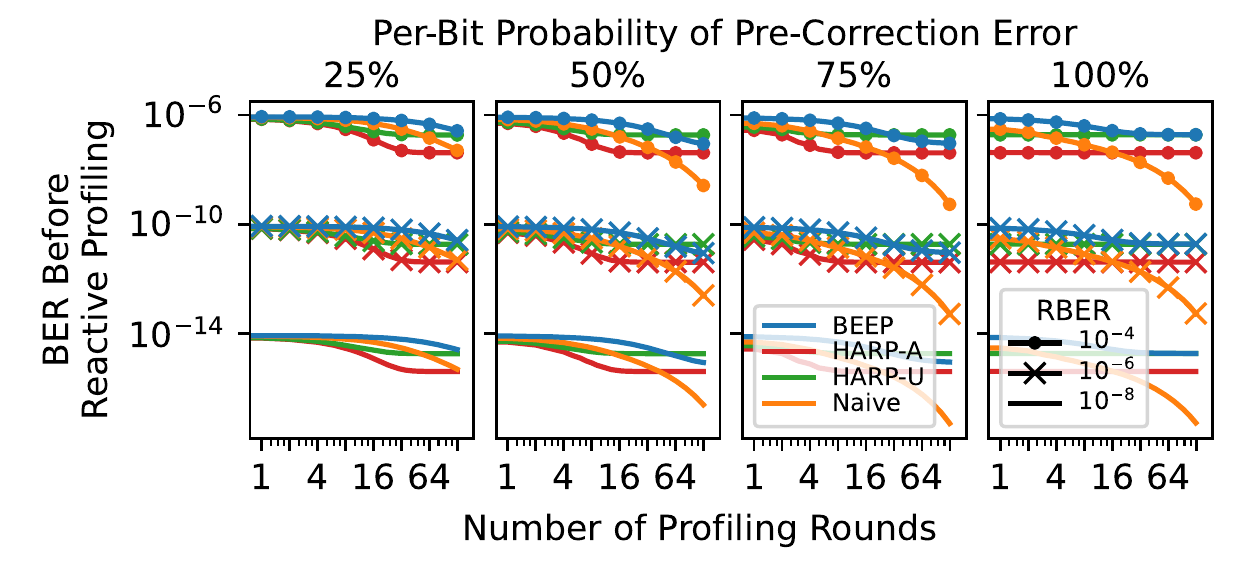}
    \end{subfigure}
    \hfill
    \begin{subfigure}[b]{0.49\textwidth}
        \centering
        \includegraphics[width=\textwidth]{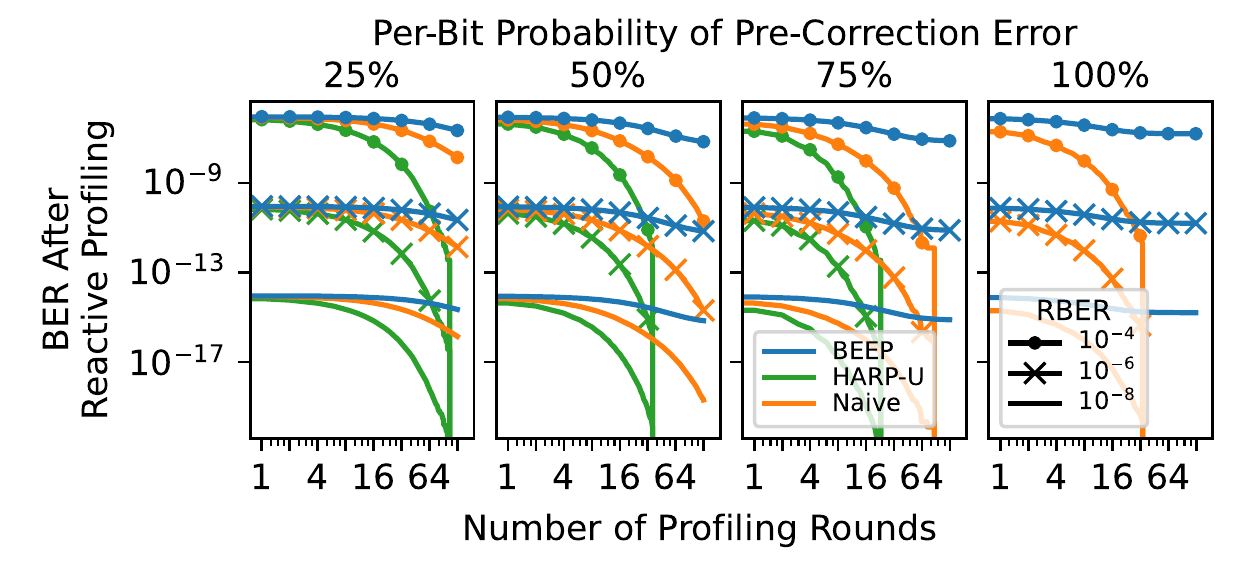}
    \end{subfigure}    
    \caption{\mps{Data-retention bit error rate (BER) using an ideal repair
    mechanism} before (left) and after (right) applying the secondary ECC.}
    \label{fig:case_study_ber}
\end{figure*}

\mpq{Fig.~\ref{fig:case_study_ber} illustrates the fraction of all bits that
are at risk of post-correction errors (i.e., the bit error rate) before
\gfii{(Fig.~\ref{fig:case_study_ber}, left)} and after
\gfii{(Fig.~\ref{fig:case_study_ber}, right)} secondary ECC is applied (i.e.,
before and after performing reactive profiling) given an x-axis number of active
profiling rounds. \mps{Each line marker shows a} different data-retention RBER
(e.g., \gfi{due to} operating at different refresh rates).  

We make three observations. First, all profilers \mps{in
Fig.~\ref{fig:case_study_ber} (left)} behave consistently with the coverage
analysis of \cref{subsubsec:direct_error_coverage}. HARP quickly
identifies all bits at risk of direct error, but leaves indirect errors to be
identified by reactive profiling. Both Naive and BEEP slowly explore different
combinations of pre-correction errors, with Naive steadily reducing the BER
given more profiling rounds while BEEP fails to do so.}

Second, \mps{Fig.~\ref{fig:case_study_ber} (left) shows} the benefit of HARP-A
knowing the on-die ECC function. While both HARP-U and HARP-A quickly identify
all bits at risk of direct errors, HARP-A also identifies bits at risk of
indirect errors, thereby considerably reducing the overall BER (and therefore,
the total number of bits) that \mpr{remain to be identified by reactive
profiling.}

\mpq{Third, \mps{Fig.~\ref{fig:case_study_ber} (right) shows that} both
HARP\footnote{\mpq{HARP-A is not shown \gfii{in Fig.~\ref{fig:case_study_ber}
(right)} because it exhibits \emph{identical} BER to HARP-U after applying
secondary ECC (i.e., because both profilers have identical coverage of bits at
risk of direct errors).}} and Naive reach a BER of zero after sufficiently many
\gfi{profiling} rounds, though Naive takes significantly more profiling rounds
to do so (e.g., $3.7\times$ for a per-bit pre-correction error probability of
75\%).} This behavior is consistent with the fact that both profilers eventually
achieve full coverage of bits at risk of direct error (shown in
\cref{subsubsec:direct_error_coverage}) \gfi{In contrast}, BEEP fails to
reach a zero probability value because it fails to achieve full coverage of bits
at risk of direct error. \mpq{Note that HARP-U immediately identifies all bits
at risk of direct errors in the first profiling round with a per-bit
pre-correction error probability of 100\%, so it is not visible in the rightmost
plot.} 

\mpq{We conclude that HARP effectively identifies all bits at-risk of error
faster than the baseline profilers, thereby enabling the repair mechanism to
safely operate at the evaluated raw bit error rates.} Although this case study
\mps{uses a simple data-retention error model that does not include
low-probability errors or other failure modes (discussed in
\cref{subsec:errors_and_models}), it} demonstrates (1) the importance of
a practical and effective error profiling algorithm in enabling a repair
mechanism to mitigate errors; and (2) the advantages that HARP provides in an
end-to-end setting by overcoming the error profiling challenges introduced by
on-die ECC.
\section{Related Work}
\label{sec:related_work}

\gfi{To our knowledge, this is the first work to (i) conduct  an analytical
study of how system-level error profiling interacts with on-die ECC, and (ii)
propose a bit-granularity error profiling algorithm to support memory chips with
on-die ECC.} Main memory error profiling is a long-standing and difficult
problem that prior works have tackled in many different ways. In our work, we
focus on error profiling algorithms in the context of on-die ECC. \gfi{We
briefly review related works which we already compared to in \gfii{prior} sections.} 

\noindent
\textbf{Profiling Without On-die ECC\gfi{.}} 
Prior works propose various error profiling algorithms in the context of
DRAM~\cite{venkatesan2006retention, liu2012raidr, khan2014efficacy,
khan2016case, khan2016parbor, khan2017detecting, patel2017reaper,
qureshi2015avatar, choi2020reducing, lee2017design,
bacchini2014characterization, sharifi2017online, kim2020revisiting,
frigo2020trrespass, cojocar19exploiting, patel2020bit, liu2013experimental,
cojocar2020are, mutlu2019rowhammer, chang2017understanding, lee2015adaptive,
kim2018solar, kim2019d, baek2014refresh, kim2003block, lin2012secret} and
emerging main memory technologies such as PCM and STT-RAM~\cite{zhang2012memory,
hamdioui2017test, tavana2017remap, qureshi2011pay}. To our knowledge, none of
these works identify or address the effects that on-die ECC has on error
profiling. Furthermore, many of the insights and/or solutions developed by these
works (e.g., algorithms for identifying low-probability
errors~\cite{qureshi2015avatar, khan2016parbor, patel2017reaper,
sharifi2017online, qureshi2011pay}) are complementary to our work and can be
integrated with HARP (e.g., during active profiling) to improve error coverage.

\noindent
\textbf{Profiling With On-die ECC.} To our knowledge, only two
works~\cite{patel2019understanding, patel2020bit} study how on-die ECC impacts
memory error characterization and profiling in detail. Unfortunately, neither
work identifies the challenges that on-die ECC introduces for error profiling.
Of these works, only BEEP~\cite{patel2020bit} is a profiling algorithm that
\mpu{accounts for} on-die ECC. However, BEEP focuses on reverse-engineering
\emph{pre-correction} error locations using a \mpu{slow algorithm} that we show
is not well suited for identifying bits at risk of \emph{post-correction} error
and still suffers from the three profiling challenges. In contrast, we
comprehensively study the three key challenges that on-die ECC introduces for
bit-granularity error profiling (\cref{sec:impl_on_error_prof}), which we
address by proposing and evaluating HARP.
\section{Conclusion}

\mpq{We study how on-die ECC affects memory error profiling and identify three
key challenges that it introduces: on-die ECC (1) exponentially increases the
number of at-risk bits the profiler must identify; (2) makes individual at-risk
bits more difficult to identify; and (3) interferes with commonly-used memory
data patterns. To overcome these three challenges, we introduce Hybrid
Active-Reactive Profiling (HARP), a new bit-granularity error profiling
algorithm that enables practical and effective error profiling for memory chips
that use on-die ECC. HARP exploits the key idea that on-die ECC introduces two
different sources of post-correction errors: (1) direct errors that result from
pre-correction errors within the data portion of the ECC codeword; and (2)
indirect errors that are a result of the on-die ECC correction process. If all
bits at risk of direct error are identified, the number of concurrent indirect
errors is upper-bounded by the correction capability of on-die ECC. Therefore,
HARP uses simple modifications to the on-die ECC mechanism to quickly identify
bits at risk of direct errors and relies on a secondary ECC within the memory
controller to safely identify indirect errors. Our evaluations show that HARP
achieves full coverage of all at-risk bits in memory chips that use on-die ECC
faster than prior approaches to error profiling. We hope that the studies,
analyses, and ideas we provide in this work will enable researchers and
practitioners alike to think about \gfii{and overcome the challenge of} how to
handle error detection and correction across the hardware-software stack in the
presence of on-die ECC.}

\begin{acks}
We thank the anonymous reviewers \gfi{of MICRO 2021} for their feedback and the
SAFARI \gfi{R}esearch \gfi{G}roup members for their feedback and the stimulating
intellectual environment they provide. \gfi{We acknowledge the generous gift
\gfii{funding} provided by our industrial partners: Google, Huawei, Intel,
Microsoft, and VMware.}
\end{acks}

\clearpage
\balance
\bibliographystyle{ACM-Reference-Format}
\bibliography{references}

\clearpage
%
%
%
%
%

\appendix

\let\underscore\_
\renewcommand{\_}{\discretionary{\underscore}{}{\underscore}}

\section{Artifact Description Appendix}

\subsection{Abstract}

Our artifacts provide the source code needed to replicate all experiments in the
paper, including all figures. The artifacts comprise two parts: (1) Monte-Carlo
simulations that generate raw data; and (2) data analysis scripts to parse,
aggregate, and plot the data. This appendix describes both parts and how to run
them to replicate our experiments. 

\subsection{Artifact \gfii{C}heck-list (\gfii{M}eta-information)}

\begin{table}[H]
  \small
  \renewcommand{\arraycolsep}{2pt}
  \begin{tabular}{p{3.2cm}p{4.6cm}}
    \textbf{Parameter} & \textbf{Value} \\\hline
    Program & C++ program and Python3/shell scripts \\\hline
    Compilation & C++11 compiler (tested with GCC 8) \\\hline
    Run-time environment & Debian 10 (or similar) Linux \\\hline
    Hardware & Standard linux system ($>=40$ GB RAM recommended for the analysis scripts) \\\hline
    Output & Plaintext data files and PDF files containing plots \\\hline
    Experiment workflow & Run simulations, aggregate output files, and run analysis scripts on the outputs \\\hline
    Experiment customization & Number of simulation jobs (i.e., Monte-Carlo samples) \\\hline
    Disk space requirement & $\approx$10-100 GB \\\hline
    Workflow preparation time & $\approx$1 day \\\hline
    Experiment completion time & $\approx$1-2 weeks  \\\hline
    Publicly available? & Zenodo~\cite{harp-artifacts} and GitHub~\cite{harpgithub} \\\hline
    Code licenses & MIT \\\hline
  \end{tabular}
\end{table}

\subsection{Description}

\subsubsection{How to Access.}

The artifacts are available on Zenodo with DOI
10.5281/zenodo.5148592~\cite{harp-artifacts} and on GitHub~\cite{harpgithub}.

\subsubsection{Hardware Dependencies.}

The artifacts are designed to run on a typical Linux system. As \cref{subsec:customization} discusses in detail, the C++ simulations are
highly parallelziable with low memory requirements, so a system with more CPUs
will reduce the overall simulation time required. In contrast, the analysis
scripts run as single tasks that consume memory in proportion to the amount of
simulation data (up to 40 GB of memory for the full evaluation configuration
described in \cref{subsec:customization}). Therefore a single machine
with a large memory may be necessary. 

\subsubsection{Software Dependencies.}

\begin{itemize}
  \item GNU Make
  \item C++11 build toolchain (tested with GCC 8 on Debian 10)
  \item Python 3~\cite{rossum2009python} with \texttt{matplotlib}~\cite{hunter2007matplotlib}, \texttt{scipy}~\cite{virtanen2020scipy}, \texttt{numpy}~\cite{harris2020array}
\end{itemize}

\subsection{Installation}

No system installation is required; the C++ code can all be built and run in
place. The C++ application depends on both Z3~\cite{de2008z3} and
EINSim~\cite{eccsimgithub}, and we have: \mpp{(1) included copies of both source
distributions in the Zenodo distribution for sake of convenience; and (2)
integrated them in the GitHub source as submodules.} However, the user may also
obtain their source files directly from the projects' repositories. Both tools
can be built using standard C++ toolchains\footnote{Note that C++17 is required
to build the latest version of Z3.} via their respective \texttt{Makefile}
projects. For convenience, we provide scripts \texttt{lib/build\_z3.sh} and
\texttt{lib/build\_einsim.sh} that the user may refer to for building the
dependencies in-place.

The primary C++ application, called \texttt{harp-artifacts}, is built using the
top-level \texttt{Makefile} project. After building Z3 within the subdirectory
\texttt{lib/z3} (which installs Z3's C++ headers and library within that
directory), the top-level \texttt{Makefile} will build \texttt{harp-artifacts}
within the top-level directory.

\subsection{Experiment \gfii{W}orkflow}
\label{sec:exp_workflow}

The artifacts are used to run three different experiments:
Fig.~\ref{fig:wasted_cap}, Fig.~\ref{fig:pre_and_post_errors}, and
Figs.~\ref{fig:eval_type1_coverage}-\ref{fig:case_study_ber}. We have provided
example scripts under \texttt{evaluations/} for automatically running the latter
two; however, we recommend the reader to extend these scripts based on their own
compute environment in order to parallelize the simulation tasks (discussed in \cref{subsec:customization}).

\subsubsection{Fig.~\ref{fig:wasted_cap}: Motivational Data.}

This is a standalone Python script that runs in $\approx$10 seconds with 120 MB
of memory using an Intel i7-7700HQ CPU @ 2.80GHz. The script replicates
Fig.~\ref{fig:wasted_cap} in a PDF output file (or can generate an interactive
Matplotlib figure per command-line arguments).

\subsubsection{Fig.~\ref{fig:pre_and_post_errors}: Post-correction Probability.}
\label{subsubsec:exp_fig_4}

This experiment comprises two steps: (1) data generation from C++ Monte-Carlo
simulations and (2) data analysis using Python scripts.

\noindent
\textbf{Step 1: Data generation.} \texttt{harp-artifacts} simulates how
different representative ECC functions (i.e., single-error correcting Hamming
codes with randomly-generated parity-check matrices) affect ECC words that
exhibit uniform-randomly generated pre-correction error patterns. Each simulated
error pattern comprises a single Monte-Carlo simulation sample. The Python
scripts then aggregate these samples as part of Step 2.

The \texttt{harp-artifacts} binary takes several command-line arguments that are
used to configure the experiment:

\begin{itemize}
  \item \emph{Analysis}: The experiment to run, either \texttt{probabilities}
  for the Fig.~\ref{fig:pre_and_post_errors} experiment or
  \texttt{evaluations} for the Figs.~\ref{fig:eval_type1_coverage}-\ref{fig:case_study_ber} experiment.
  \item \emph{ECC dataword length ($k$)}: The length (in bits) of a single-error
  correcting Hamming code dataword. This parameter defines the type of ECC code
  that will be generated and simulated (i.e., its generator and parity-check
  matrix dimensions). Our evaluations are all shown with $k=64$, though we
  verified that all results and conclusions are consistent with other
  representative values, such as $k=128$ (\cref{subsubsec:sim_strat}).
  \item \emph{Number of ECC codes}: The number of Hamming code instances to
  generate and simulate. Each code's generator and parity-check matrices are
  created uniform-randomly according to the random seed command-line parameter.
  \item \emph{Number of ECC words}: The number of ECC words to simulate for each
  randomly-generated ECC code.
  \item \emph{Random seed}: The random seed to use for the first ECC code
  generated. The random seed is then incremented when generating each subsequent
  ECC code. 
\end{itemize}

\noindent
To run the simulations for this experiment, the \texttt{analysis} parameter
should be provided as \texttt{probabilities}. \cref{subsec:customization}
summarizes different configurations for the remaining parameters and their
expected runtime and memory usage impact. All simulation output will be given on
\texttt{stdout}, which must be redirected to a text file to pass the data to the
Python analysis scripts in the next step.

\noindent
\textbf{Step 2: Data analysis.} The Python script
\texttt{script/figure\_4-parse\_postcorrection\_probabilities\_data.py} 
accepts an input file containing the raw output from the previous step. The
script will then parse, aggregate, analyze, and plot the data using
\texttt{matplotlib} (either interactively or saved to a PDF file, based on a
command-line switch).

\subsubsection{Figs.~\ref{fig:eval_type1_coverage}-\ref{fig:case_study_ber}:
Profiler Evaluation.}

This experiment runs in two parts: (1) data generation from C++ Monte-Carlo
simulations and (2) data analysis using Python scripts. The experiment workflow
is nearly identical to that of \cref{subsubsec:exp_fig_4}.

\noindent
\textbf{Step 1: Data generation.} All \texttt{harp-artifacts} command-line
parameters operate the same way as in \cref{subsubsec:exp_fig_4}, except the
\texttt{analysis} parameter must be given as \texttt{evaluations} for this
experiment.

In \texttt{evaluations} mode, \texttt{harp-artifacts} simulates the coverage
achieved by the different profiling mechanisms that we evaluate in
\cref{sec:evaluations}. These simulations are extremely time consuming
due to the complexity of calculating error coverage, which requires a large
number of computations, including repeated SAT solver invocations.
\cref{subsec:customization} provides the expected runtimes for different
configurations that the reader may use to estimate a viable configuration based
on their available compute resources.

\noindent
\textbf{Step 2: Data analysis.} The Python script
\texttt{script/figures\_6to10-parse\_evaluation\_data.py} accepts an input file
containing the raw output from the previous step and is run the same way as the
script in \cref{subsubsec:exp_fig_4}. Each figure is output either interactively
or in individual PDF files based on a command line switch.

\subsection{Evaluation and \gfii{E}xpected \gfii{R}esults}
\label{subsec:eval_and_expected_results}

To replicate the results in this paper, it suffices to run each of the
experiments as discussed in \cref{sec:exp_workflow}. Note that, as we
discuss in \cref{subsec:customization}, our full evaluation configuration
is long-running due to the large number of Monte-Carlo samples that we simulate.
It is \emph{not} necessary to simulate this many samples to replicate our
results; even with relatively few samples, the data yields similar conclusions.
In general, we leave the reader to determine how many samples they can
realistically simulate based on their available compute resources. To facilitate
this, \cref{subsec:customization} provides runtime estimates measured
during our own evaluations.

\subsection{Experiment \gfii{C}ustomization}
\label{subsec:customization}

As \cref{subsubsec:exp_fig_4} describes, \texttt{harp-artifacts} has
independent command-line parameters to control the number of ECC codes and ECC
words simulated. Using the ECC code parameter, it is possible to parallelize the
simulations across different invocations of \texttt{harp-artifacts} (e.g., as
independent jobs on different compute nodes). For example, simulating 100 ECC
words for each of 1000 ECC codes can be done by simulating 10 ECC codes per job
across 100 independent jobs (and incrementing the random seed by 10 for each
subsequent job to avoid repeating the same experiment). The Python data analysis
scripts will aggregate the raw data, regardless of how the ECC codes are
partitioned.

\subsection{Estimating Runtime and Memory Usage}
\label{subsec:estimating}

Running the experiments for Fig.~\ref{fig:pre_and_post_errors}, and
Figs.~\ref{fig:eval_type1_coverage}-\ref{fig:case_study_ber} requires
accounting for available compute and memory resources. In this section, we
discuss different configurations of \texttt{harp-artifacts} and how they impact
runtime and memory usage when running on a compute cluster comprising Intel Xeon
Gold 5118 systems.

\subsubsection{Fig.~\ref{fig:pre_and_post_errors} Estimation.}

Table~\ref{tab:eval_estimate_fig4} summarizes the expected runtime and disk
usage of a \emph{single task} for different \texttt{harp-artifacts}
configurations when running the Fig.~\ref{fig:pre_and_post_errors} experiment.
In the evaluations we present in \cref{sec:evaluations}, we run 16
instances of the ``Evaluated'' configuration shown in
Table~\ref{tab:eval_estimate_fig4}, amounting to a total of 74 GB of disk usage
and approximately 1 CPU-day of total execution time. In general, the
Fig.~\ref{fig:pre_and_post_errors} simulations dump a large amount of raw
data. However, the runtime and memory usage are relatively low, so the reader
may feasibly replicate our full evaluation configuration.

\begin{table}[H]
  \small
  \renewcommand\tabcolsep{3pt}
  \centering
  \begin{tabular}{llllll}
    \textbf{Configuration} & \textbf{K} & \textbf{ECC codes} & \textbf{ECC words} & \textbf{Runtime} & \textbf{Disk Usage} \\\hline
    Reduced & 64 & 10 & 100 & $\approx$30 sec. & 4.8 MB \\\hline
    Evaluated & 64 & 100 & 10000 & $\approx$90 min. & 4.6 GB \\\hline
  \end{tabular}
  \caption{Estimated runtime and disk usage for one instance of
  \texttt{harp-artifacts} when running the Fig.~\ref{fig:pre_and_post_errors} experiment.}
  \label{tab:eval_estimate_fig4}
  \vspace{-2em}
\end{table}

\subsubsection{Figs.~\ref{fig:eval_type1_coverage}-\ref{fig:case_study_ber} Estimation.}
\label{subsubsec:eval_runtime_estimate}

Table~\ref{tab:eval_estimate_fig5_10} summarizes the expected runtime and disk
usage of a \emph{single task} for different \texttt{harp-artifacts}
configurations when running the
Figs.~\ref{fig:eval_type1_coverage}-\ref{fig:case_study_ber} experiments. In
the evaluations we present in \cref{sec:evaluations}, we run 256
instances of the ``Evaluated'' configuration shown in
Table~\ref{tab:eval_estimate_fig5_10}, amounting to a total of 3.4 GB of disk
usage and approximately 14 CPU-years of total execution time. In general, the
memory and disk usage for \texttt{harp-artifacts} is negligible in this
experiment, but the runtime can be a considerable limitation.

\begin{table}[H]
  \small
  \centering
  \renewcommand\tabcolsep{2.8pt}
  \begin{tabular}{llllll}
    \textbf{Configuration} & \textbf{K} & \textbf{ECC codes} & \textbf{ECC words} & \textbf{Runtime} & \textbf{Disk Usage} \\\hline
    Reduced & 8 & 1 & 1 & $\approx$10 sec. & - \\\hline
    Reduced & 16 & 1 & 1 & $\approx$30 sec. & - \\\hline
    Reduced & 32 & 1 & 1 & $\approx$5 min. & - \\\hline
    Reduced & 64 & 1 & 1 & $\approx$30 min. & - \\\hline
    Evaluated & 64 & 10 & 100 & $\approx$20 days & 13 MB \\\hline
  \end{tabular}
  \caption{Estimated runtime and disk usage for one instance of
  \texttt{harp-artifacts} when running the Figs.~\ref{fig:eval_type1_coverage}-\ref{fig:case_study_ber} experiment.}
  \label{tab:eval_estimate_fig5_10}
  \vspace{-2em}
\end{table}

As we mention in \cref{subsec:eval_and_expected_results}, it is not
necessary to run the full ``Evaluated'' configuration to replicate our results.
By the nature of Monte-Carlo simulation, simulating more ECC codes and ECC words
improves the accuracy of the final estimates. However, the end conclusions are
already apparent with relatively few samples (albeit with more noise in the
data). Therefore, if running the full ``Evaluated'' configuration is infeasible,
we recommend running a reduced configuration based on the available compute
resources. Note that the average runtime scales linearly with the number of
samples because the same computation is performed for each sample. For example,
reducing the ``Evaluated'' configuration to 1 ECC code per task would reduce the
average runtime by a factor of 10, resulting in jobs that comple in
approximately 48 hours. 

We find that the Python analysis for the full evaluated configuration (i.e., all
3.4 GB of output) takes approximately 3 hours and consumes 40 GB of memory. We
observe that the memory usage is roughly linear in the number of evaluated ECC
codes and words.

\end{document}